\documentclass[journal]{IEEEtran}

\usepackage{amsmath,amsfonts}
\usepackage[ruled,linesnumbered]{algorithm2e}
\usepackage{array}
\usepackage{subfig}
\usepackage{multirow}
\usepackage{graphicx}
\usepackage{caption}
\usepackage{textcomp}
\usepackage{url}
\usepackage{verbatim}
\usepackage{graphicx}
\usepackage{mathtools}
\usepackage{algorithmic}
\usepackage{cite}
\usepackage{booktabs} 
\usepackage{mathrsfs}
\usepackage{tikz}
\begin{document}
\pagestyle{empty}  % no page number for the second and the later pages

\title{Leveraging Self-Supervised Learning for MIMO-\\OFDM Channel Representation and Generation}

\author{Zongxi~Liu,~\IEEEmembership{Member,~IEEE,}
        Jiacheng~Chen,~\IEEEmembership{Member,~IEEE,}
        Yunting~Xu,~\IEEEmembership{Member,~IEEE,}
        Ting~Ma,~\IEEEmembership{Member,~IEEE,}
        % Bo~Qian,~\IEEEmembership{Member,~IEEE,}
        Jingbo~Liu,~\IEEEmembership{Member,~IEEE,}
        Haibo~Zhou,~\IEEEmembership{Senior Member,~IEEE,}
        and
        Dusit~Niyato,~\IEEEmembership{Fellow,~IEEE}

\thanks{Z. Liu, Y. Xu and H. Zhou (Corresponding author) are with the School of Electronic Science and Engineering, Nanjing University, Nanjing, China, 210023
(e-mail: zongxiliu@smail.nju.edu.cn, yuntingxu@smail.nju.edu.cn, haibozhou@nju.edu.cn).}
\thanks{J. Chen is with the Department of Strategic and Advanced Interdisciplinary Research, Peng Cheng Laboratory, Shenzhen 518000, China (email: chenjch02@pcl.ac.cn, boqian@pcl.ac.cn).}
\thanks{T. Ma is with the School of Electronic and Optical Engineering, Nanjing University of Science and Technology, Nanjing 210094, China (email: tingma@njust.edu.cn).}
\thanks{J. Liu is with School of Electronic Information and Electrical Engineering, Shanghai Jiao Tong University, Shanghai, China, 200240.(e-mail: liujingbo@sjtu.edu.cn).}
\thanks{D. Niyato is with the School of Computer Science and Engineering, Nanyang Technological University, Singapore, 639798 (e-mail: dniyato@ntu.edu.sg).
}
}


\maketitle
\thispagestyle{empty} 
\begin{abstract}
In communications theory, the capacity of multiple input multiple output-orthogonal frequency division multiplexing (MIMO-OFDM) systems is fundamentally determined by wireless channels, which exhibit both diversity and correlation in spatial, frequency and temporal domains. It is further envisioned to exploit the inherent nature of channels, namely representation,  to achieve geolocation-based MIMO transmission for 6G, exemplified by the fully-decoupled radio access network (FD-RAN). Accordingly, this paper first employs self-supervised learning to obtain channel representation from unlabeled channel, then proposes a channel generation assisted approach for determining MIMO precoding matrix solely based on geolocation. Specifically, we exploit the small-scale temporal domain variations of channels at a fixed geolocation, and design an ingenious pretext task tailored for contrastive learning. Then, a Transformer-based encoder is trained to output channel representations. We further develop a conditional diffusion generator to generate channel representations from geolocation. Finally, a Transformer-encoder-based decoder is utilized to reconstruct channels from generated representations, where the optimal channel is selected for calculating the precoding matrix for both single and dual BS transmission. We conduct experiments on a public ray-tracing channel dataset, and the extensive simulation results demonstrate the effectiveness of our channel representation method, and also showcase the performance improvement in geolocation-based MIMO transmission.
\end{abstract}


\section{Introduction}
\IEEEPARstart{I}{n} wireless communications, the channel characterizes signal propagation in the physical environment and determines the quality of transmission. In multiple input multiple output-orthogonal frequency division multiplexing (MIMO-OFDM) systems, the channel is typically represented in the spatial, frequency, and temporal domains\cite{tse2005fundamentals,proakis2008digital,wang2023complete}. The spatial domain describes the scattering of multipath signals over the angular domain, with respect to the angle of departure (AoD) at the transmitter and the angle of arrival (AoA) at the receiver. The frequency domain characterizes the scattering of multipath fading over the delay domain, capturing the frequency-selective fading of the channel. And the temporal domain reflects the environmental dynamics in terms of the multipath effect. Here, we refer to the channel at a fixed geographic location, thereby excluding the Doppler effect. Through studying the channel, its characteristics  can be effectively utilized to facilitate MIMO transmission, such as beamforming or spatial multiplexing.

Furthermore, the channel exhibits correlation in spatial, frequency and temporal domains. With uniform linear array (ULA) or uniform planar array (UPA) antennas, electromagnetic waves arrive at different antennas with a fixed phase difference. Additionally, there is a fixed interval between different carriers, resulting in a fixed phase difference among the channels on different subcarriers. At a specific location, the channel exhibits temporal correlation as the predominant electromagnetic waves follow approximately the same propagation path. The correlation implies that channel can be represented in a form with higher information entropy, namely channel representation, which is critical to understand the inherent characteristics of channel.

Currently, there is still a lack of efficient methods to obtain channel representation from the raw channel data. To this end, we attempt to employ self-supervised learning on the 
unlabeled channel dataset to obtain channel representations. In the field of artificial intelligence (AI), self-supervised learning is emerging as a dominant approach, driven by observations that models trained through unsupervised learning on large datasets often outperform those trained with supervised learning on smaller datasets. In both computer vision (CV) and natural language processing (NLP), it has become a standard practice to employ self-supervised learning methods to extract input representations (such as word vectors in NLP or image features in CV) and then utilize these representations in downstream tasks\cite{devlin2018bert, radford2018improving,brown2020language,radford2019language, he2022masked,he2020momentum,chen2020simple,grill2020bootstrap}. Similarly, in this paper, we learn the channel representations first, and then evaluate the representations via carefully-designed downstream MIMO transmission tasks. 

However, applying self-supervised learning to channels presents several challenges. First, there is no applicable pretext task tailored for channels, as the correlation of channels differs fundamentally from the contextual correlation in NLP or the pixel correlation in CV. Thus, a suitable pretext task must be specifically designed based on the unique characteristics of the channel. Second, unlike in NLP or CV, where established metrics such as accuracy for sentiment analysis or mean average precision for object detection are commonly used to assess the quality of representations\cite{pang2002thumbs, ren2015faster}. There is no standard metric for evaluating channel representations. Consequently, we need to develop an appropriate metric to measure the effectiveness of a channel representation. Moreover, since channels, unlike images, do not exhibit the translation invariance that characterizes the inductive bias of convolutional neural network (CNN), CNN-based models may not be suitable for channel feature extraction\cite{chen2023viewing}. Thus, a non-CNN-based network must be developed specifically for the self-supervised learning of channels.

To solve the aforementioned challenges, this paper proposes a complete pipeline for representation and generation of MIMO-OFDM channels. Initially, channels from different geolocations are utilized as negative samples, while channels from the same geolocations at various time instants serve as positive samples. This novel setup facilitates the design of the pretext task for contrastive learning, which enables the development of an encoder capable of mapping channel matrices into channel representations in latent space. To avoid the inductive bias inherent in CNNs, this encoder employs a Transformer-encoder-based architecture. The Transformer-encoder has demonstrated its efficacy as a robust feature extractor, yielding impressive outcomes in both CV and NLP \cite{dosovitskiy2020image, devlin2018bert, vaswani2017attention}. Due to its lack of inductive bias, the Transformer-encoder can outperform CNNs when provided with sufficiently large datasets.

Furthermore, to assess the efficacy of the obtained representations, we introduce a geolocation-based MIMO transmission as the downstream task and use its throughput as the evaluation metric. The geolocation-based MIMO transmission task utilizes solely geolocation information to determine MIMO transmission parameters (like precoding matrix), i.e. without relying on channel state information (CSI) feedback or channel reciprocity. Considering the strong correlation between transmission parameters and the inherent nature of channels, selecting the optimal precoding matrix needs to consider factors such as AoD. Thus, this task serves as an effective metric of channel characterization. Another important reason for selecting the geolocation-based MIMO transmission task is its potential application in 6G scenarios, including fully-decoupled radio access network (FD-RAN) \cite{yu2019fully,qian2023enabling} and integrated sensing and communication\cite{zeng2021toward}. Specifically, in the FD-RAN scenario, a conventional base station (BS) is split into an uplink BS, a downlink BS, and a control BS. By means of hardware separation, uplink and downlink transmissions are completely independent, and resource allocation can be flexibly adjusted to meet the diverse and changing service and coverage requirements of user equipment (UE). However, the decoupling of BS also indicates that traditional CSI feedback is no longer applicable. Thus, feedback-free MIMO transmission approaches are required\cite{liu2023enabling,liu2023leveraging,zhao2023channel}.

To achieve geolocation-based MIMO transmission, we train a conditional diffusion generator that incorporates geolocation information as the condition and generates representations in latent space. On the one hand, the generator functions as a mapper that transforms geolocation information into representations. On the other hand, leveraging the learned distribution of these representations, the generator can repeatedly generate representations, from which the optimal one is selected. Subsequently, a decoder is trained to reconstruct this optimal representation into a channel, which is then utilized for the geolocation-based MIMO transmission task. The decoder, also based on the Transformer-encoder, maintains symmetry with the encoder's structure and simultaneously avoids the checkerboard effect commonly associated with deconvolution operation. We summarize the main contributions of this paper across the following three aspects:
\begin{itemize}
\item We propose a self-supervised learning approach aiming at obtaining the representation of the wireless channel. By exploiting the small-scale temporal domain variations of channels, a pretext task for contrastive learning is designed to train an encoder that can map channel matrices to representations in latent space. Additionally, our encoder are designed using the Transformer-encoder to circumvent the issues associated with the inductive bias found in CNNs.
\item To evaluate the channel representation,  we utilize the downstream task of geolocation-based MIMO transmission, which can capture the inherent nature of channel. A conditional diffusion generator is proposed to  generate channel representations first, then a decoder is used to remap the representations to the channel matrices. The generator can be utilized repeatedly in conjunction with a decoder to derive optimal representations and channels for geolocation-based MIMO transmission.
\item We conduct extensive simulations on a public ray-tracing-based channel dataset \cite{alkhateeb2019deepmimo} and demonstrate that our approach outperforms the baseline and other existing methods in both the uniform manifold approximation and projection (UMAP) visualizations and the geolocation-based MIMO transmission task.
\end{itemize}

The remainder of this paper is structured as follows.
Section II introduces the related work. Section III presents  the system model and the downstream MIMO transmission task.
Section IV describes the proposed solution, followed by the simulation results given in Section V.
Finally, we conclude the paper in Section VI. 

\section{Related Works}
\subsection{Self-supervised Learning}
Self-supervised learning has evolved into two main approaches: context-based and contrastive-based self-supervised learning. The fundamental principle of context-based self-supervised learning is removing part of the information and allowing the model to recover it through context. Examples include cloze tasks in NLP, as well as inpainting and video frame interpolation in computer vision. \cite{mikolov2013efficient, devlin2018bert, pathak2016context, he2022masked,sermanet2018time}. Contrastive-based supervised learning, on the other hand, develops representations by constructing positive and negative samples and analyzing their similarities and differences, with instance discrimination being the most classic task\cite{wu2018unsupervised, he2020momentum,chen2020simple}. For the channel matrix, frequency and spatial domain correlation are explicit, whereas temporal correlation is implicit. In other words, the time-domain variation of the channel at a fixed location cannot be modeled as a time series, namely context, due to random environmental variations. However, this time-domain variation of the channel can be treated as positive sample pairs in contrastive learning. Therefore, we adopt contrastive learning as a framework for channel self-supervised learning and design a pretext task suitable for the channel matrix.
\subsection{Self-supervised Learning for Channel}
 Due to the substantial overhead of CSI feedback in massive MIMO scenarios, the compression-recovery task of CSI has emerged as a significant research focus. Consequently, there has been progress in applying self-supervised learning to channel analysis \cite{deng2022supervised,chen2023viewing,wen2018deep}. For instance, \cite{wen2018deep} proposed CsiNet, which utilizes the concept of an autoencoder to compress and recover channel matrices in the angular domain, with the recovery performance evaluated using the normalized mean squared error (NMSE). \cite{chen2023viewing} also employed an autoencoder structure for self-supervised learning but opts for long short-term memory networks instead of CNNs, arguing that CNN's inductive bias of translation invariance is not applicable to channel matrices. \cite{deng2022supervised} adopted a contrastive learning approach to extract channel representation, treating geographically neighboring channels as positive samples and non-neighboring channels as negative samples. Additionally, a positioning task was proposed to evaluate the quality of the representations. Different from the works
mentioned above, our work introduces the temporal domain to the analysis of the channel matrix, an aspect not addressed in the aforementioned studies. Our works consider channels from different geolocations as negative samples, while channels from the same geolocations at various time instants serve as positive samples. Furthermore, for the downstream task, we utilize a generator and employ a decoder based on a Transformer-encoder module instead of CNN, effectively bypassing the influence of the CNN’s inductive bias.
\subsection{Generative Model for Channel Matrix}
Generative models have demonstrated remarkable performance across various domains, prompting several studies to explore channel generation using these models\cite{liu2023enabling, yang2019generative, sengupta2023generative}. \cite{liu2023enabling} initially trained a variational autoencoder network to generate the optimal precoding matrix for the channel. Subsequently, channels at the same geographic location but at different times were characterized in the latent space, where appropriate parameters are chosen to generate a representative precoding matrix for the point. This approach exemplifies the use of generative models to enhance transmission performance. \cite{yang2019generative} employed generative adversarial networks to generate channels, using the generator and discriminator to closely approximate the real captured channel. \cite{sengupta2023generative} utilized denoising diffusion probabilistic models to generate channels and achieves superior results compared to generative adversarial networks. The evaluation metric employed was the channel power delay profile similarity. In contrast to the aforementioned studies, our approach generates the channel representation instead of the channel matrix itself. This is because the latent space has greater information entropy, allowing the generative model to exhibit enhanced creativity. Furthermore, we employ a decoder to generate the channel matrix from the generated representation, facilitating the geoloaction-based MIMO transmission task.

\section{System Model}
In this section, we describe the mathematical characterization of channels and the format of the dataset used to train the neural network. Furthermore, we present a scenario of geolocation-based MIMO transmission under the framework of FD-RAN and formulate two tasks in this scenario to serve as evaluation metrics for the channel representation.

\subsection{Mathematical Characterization of Channel}
\begin{figure}[htbp]
\centerline{\includegraphics[width=0.9\linewidth]{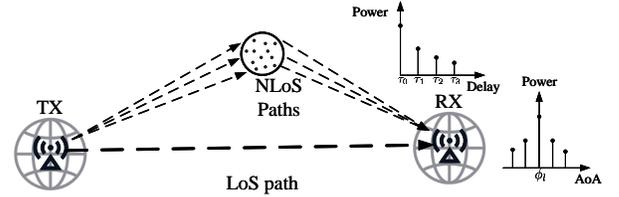}}
\caption{The mathematical characterization of channels that account for the distribution of AoA and delays.}
\label{fig:channel}
\end{figure}
Consider that there exist $ N_b$ BSs and $ N_u$ user equipments (UEs). Each BS is equipped with $N_T=N_xN_yN_z$ 3-Dimension array antennas, where $N_x, N_y, N_z$ represent the number of antennas in the $x$, $y$ and $z$ directions, respectively. Each user has $N_{R}$ uniform linear array antennas, and the distance between adjacent antennas is $d$. The OFDM technique is utilized for transmission, and the downlink channel between the BS $b \in \left\{1,2,\cdots,N_b\right\}$ and the UE $u \in \left\{1,2,\cdots,N_u\right\}$ can be expressed as follows for time $t$ and subcarrier $k$:
\begin{equation}\label{eq:channel model}
\begin{split}
\mathbf{H}^{b,u,t, k}=\sum_{\ell=1}^{N_p} \sqrt{\frac{\Gamma_{\ell}^{b,u,t}}{N_k}} e^{j\left(\vartheta_{\ell}^{b,u,t}+\frac{2 \pi k}{N_k} \tau_{\ell}^{b,u,t} B\right)}\\ \mathbf{a}_{UE}\left(\theta^{b,u,t}\right)\mathbf{a}_{BS}^H\left(\phi_{\mathrm{az}}^{b,u,t}, \phi_{\mathrm{el}}^{b,u,t}\right)\in \mathbb{C}^{N_{R}\times N_T},
\end{split}
\end{equation}
where $N_k$ and $N_p$ denote the total number of subcarriers and multipaths, respectively. $B$ represents the bandwidth. It is important to note that in the real world, $N_p \rightarrow \infty$. As a result, $\mathbf{H}^{b,u,t, k}$ is difficult to model accurately. As shown in the Fig. \ref{fig:channel}, $\Gamma_{\ell}^{b,u,t}$, $\vartheta_{\ell}^{b,u,t}$ and $\tau_{\ell}^{b,u,t}$ are the power, phase and delay of the $\ell$-th path, respectively. $\theta^{b,u,t}$ denotes the AoAs at the UE, as well as $\phi_{\text{az}}^{b,u,t}$ and $\phi_{\text{el}}^{b,u,t}$ represent the azimuth and elevation AoDs from the BS, respectively. $\mathbf{a}_{UE}\left(\theta^{b,u,t}\right)$ and $\mathbf{a}_{BS}\left(\phi_{\mathrm{az}}^{b,u,t}, \phi_{\mathrm{el}}^{b,u,t}\right)$ denote the array response vectors of the UE and BS, which are defined as
\begin{equation}\label{UE Impulse Response}
\begin{split}
\mathbf{a}_{UE}\left(\theta^{b,u,t}\right)  = 
\left[1, e^{j k d \cos (\theta^{b,u,t})},  \ldots, e^{j k d (N_{R}-1) \cos (\theta^{b,u,t})}\right]_{,}^T
\end{split}
\end{equation}
\begin{equation}\label{BS Impulse Response}
\begin{split}
\mathbf{a}_{BS}\left(\phi_{\mathrm{az}}^{b,u,t}, \phi_{\mathrm{el}}^{b,u,t}\right)= \mathbf{a}_z\left(\phi_{\mathrm{el}}^{b,u,t}\right)\otimes
\mathbf{a}_y\left(\phi_{\mathrm{az}}^{b,u,t}, \phi_{\mathrm{el}}^{b,u,t}\right) \\\otimes \mathbf{a}_x\left(\phi_{\mathrm{az}}^{b,u,t}, \phi_{\mathrm{el}}^{b,u,t}\right),
\end{split}
\end{equation}
where $\otimes$ denotes the kronecker product. $\mathbf{a}_x\left(\cdot\right)$, $\mathbf{a}_y\left(\cdot\right)$, $\mathbf{a}_z\left(\cdot\right)$ are the array impulse response vectors of BS in the $x$, $y$ and $z$ directions, which can be expressed as
\begin{equation}\label{xyz Impulse Response}
\begin{aligned}
\mathbf{a}_x\left(\phi_{\mathrm{az}}^{b,u,t}, \phi_{\mathrm{el}}^{b,u,t}\right)=&\left[1, e^{j k d \sin \left(\phi_{\mathrm{el}}^{b,u,t}\right) \cos \left(\phi_{\mathrm{az}}^{b,u,t}\right)}, \right.\\&\left.
\ldots, e^{j k d\left(N_x-1\right) \sin \left(\phi_{\mathrm{el}}^{b,u,t}\right) \cos \left(\phi_{\mathrm{az}}^{b,u,t}\right)} \right] ^T\\\mathbf{a}_y\left(\phi_{\mathrm{az}}^{b,u,t}, \phi_{\mathrm{el}}^{b,u,t}\right)=&\left[1, e^{j k d \sin \left(\phi_{\mathrm{el}}^{b,u,t}\right) \sin \left(\phi_{\mathrm{az}}^{b,u,t}\right)}, \right.\\&\left.
\ldots, e^{j k d\left(N_y-1\right) \sin \left(\phi_{\mathrm{el}}^{b,u,t}\right) \sin \left(\phi_{\mathrm{az}}^{b,u,t}\right)} \right] ^T,\\
\mathbf{a}_z\left(\phi_{\mathrm{el}}^{b,u,t}\right)\!\!=\!\left[1,\vphantom{e^{j k d \cos \left(\phi_{\mathrm{el}}^{b,u,t}\right)}} e\right.&\left.^{\!\!j k d \cos \left(\phi_{\mathrm{el}}^{b,u,t}\right)}, \!\ldots\!, e^{j k d\left(N_z-1\right) \cos \left(\phi_{\mathrm{el}}^{b,u,t}\right)}\right]_{.}^T
\end{aligned}
\end{equation}

At time $t$, the downlink channel between the BS $b$ and the UE $u$ can be expressed as $\mathbf{H}^{b,u,t} = \bigcup_{k=1}^{N_k}\mathbf{H}^{b,u,t,k} \in \mathbb{C}^{N_k\times N_{R}\times N_T}$, where $\bigcup$ represents the operation of concatenating elements along a new dimension. We collect $N_t$ channels between this pair of BS and user and denote them as $\mathbf{H}^{b,u} = \bigcup_{t=1}^{N_t}\mathbf{H}^{b,u,t}\in \mathbb{C}^{N_t\times N_k\times N_{R}\times N_T}$. Note that $t$ does not need to be a specific time point. 
Different values of $t$ only indicate that the channel matrices are different due to the time-varying characteristics of the channel. We use $\mathbf{loc}^{b,u} \in \mathbb{C}^{3 \times 1}$ to denote the relative geolocation between $b$ and $u$ in $x,y$ and $z$ directions. Therefore, the dataset can be expressed as $\left\{\mathbf{loc}^{b,u},\mathbf{H}^{b,u}\right\}_{ N_b\times N_u}$.
\subsection{Geolocation-based MIMO Transmission}
In a downlink MIMO-OFDM system, the signal received by the UE $u$ from $N_b$ BSs is given by
\begin{equation}\label{MIMO OFDM transmission}
\mathbf{y}^{u,t,k} = \sum_{b=1}^{ N_b}\left(\mathbf{H}^{b,u,t,k}\mathbf{W}^{b,u,k}\right)x^{b,u,t,k} + \mathbf{n}^{b,u,t,k},
\end{equation}
where $x^{b,u,t,k} \in \mathbb{C}^{N_{l}\times1}$ and $\mathbf{y}^{u,t,k} \in \mathbb{C}^{N_{R}\times1}$ are the transmitted and received signals, respectively, and $\mathbf{n}^{b,u,t,k} \in \mathbb{C}^{N_{R} \times 1} \sim \mathcal{C N}\left(0, \sigma_n^2 \cdot \mathbf{I}\right)$ is the additive zero-mean, complex-valued, white Gaussian noise with variance $\sigma_n^2$. $\mathbf{W}^{b,u,k} \in \mathbb{C}^{N_T\times N_{l}}$ is the precoding matrix to map the $N_{l}$ layers of data into $N_T$ transmit antenna ports. It is important to note that since there is no real-time CSI feedback, $\mathbf{W}^{b,u,k}$ is a time-invariant matrix. Given that different BSs send the same OFDM symbols to the UE, the signals from different BSs can be interpreted as multipath transmissions. Consequently, the inter-symbol interference caused by the multipath effect can be effectively mitigated by using the cyclic prefix technique, so that the received signal can be directly expressed as a sum of signals from multiple BSs.

\begin{figure}[htbp]
\centerline{\includegraphics[width=0.8\linewidth]{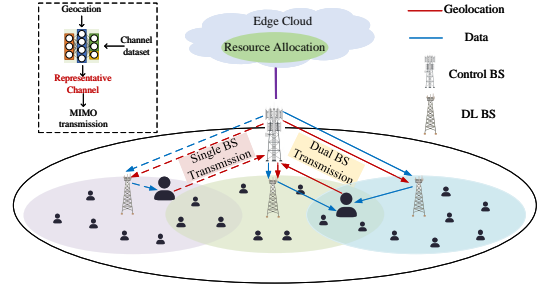}}
\caption{Illustration of single BS transmission and dual BS joint transmission in FD-RAN.}
\label{fig:FD-RAN}
\end{figure}

Next, we introduce the specific scenario and present two downstream tasks for channel representations in gelocation-based MIMO transmission. In FD-RAN, a conventional BS is divided into an uplink BS, a downlink BS, and a control BS. This hardware separation allows for completely independent uplink and downlink transmissions, enabling resource allocation to be flexibly adjusted to meet the requirements of varying and evolving services. Due to the physical separation of uplink and downlink BSs, geolocation-based MIMO transmission without CSI feedback is employed in FD-RAN. As illustrated in Fig. \ref{fig:FD-RAN}, we outline the tasks that need to be addressed in the case of single BS MIMO transmission and dual BS joint MIMO transmission with power allocation in FD-RAN, respectively.
\subsubsection{Single BS MIMO Transmission}
We consider a scenario with a single BS for MIMO transmission and no power allocation. For a given time $t$ and subcarrier 
$k$, our objective is to design a precoding matrix that maximizes throughput:
\begin{equation}\label{task1 ini obj}
    \begin{aligned}
&\max\limits_{\mathbf{W}^{b,u,t,k}}\log_2\left(1+\frac{\|\mathbf{H}^{b,u,t,k}\mathbf{W}^{b,u,t,k}\|_2^2}{\sigma_n^2}\right)
\\&\text{s.t.} \left\Vert\mathbf{W}^{b,u,t,k}\right\Vert_{F}=1,\forall b\in  \left\{1,2,\cdots,N_b\right\}, \\&\forall u\in  \left\{1,2,\cdots,N_u\right\},\forall k\in \left\{1,2,\cdots,N_k\right\},
\end{aligned}
\end{equation}

where $\|\cdot\|_2$ is the  L2 norm. \cite{marzetta1999capacity} demonstrates that the optimal precoding matrix for this problem is given by $\mathbf{W}_{*}^{b,u,t,k}=\mathscr{V}_{N_{l}}\left(\mathbf{H}^{b,u,t,k}\right)/\|\mathscr{V}_{N_{l}}\left(\mathbf{H}^{b,u,t,k}\right)\|_F$, where $\mathscr{V}_{N_{l}}\left(\cdot\right)$ represents the first $N_{l}$ columns of the unit-norm right singular vector matrix obtained from the singular value decomposition (SVD) of the channel and $\|\cdot\|_F$ is the Frobenius norm. However, since there is no CSI feedback in FD-RAN, the precoding matrix $\mathbf{W}^{b,u,t,k}$ cannot change in response to variations in the channel matrix $\mathbf{H}^{b,u,t,k}$. Therefore, we aim to identify a \textit{representative channel} $\mathbf{H}_{re}^{b,u,k} \in \mathbb{C}^{N_{R}\times N_T}$ through which the obtained optimal precoding matrix can maximize the average spectrum efficiency over time: 
\begin{equation}\label{task1 obj}
\begin{aligned}
&\mathbf{H}_{re,*}^{b,u,k} =\arg\max_{\mathbf{H}_{re}^{b,u,k}}\text{TASK1}_{N_{l}}\\
&\operatorname{s.t.} \text{TASK1}_{N_{l}} =\\
 &\frac{1}{N_kN_t}\sum_{t=1}^{N_t}\sum_{k=1}^{N_k}\log_2\left(1+\frac{\left\|\mathbf{H}^{b,u,t,k}\frac{\mathscr{V}_{N_{l}}\left(\mathbf{H}_{re}^{b,u,k}\right)}{\|\mathscr{V}_{N_{l}}\left(\mathbf{H}_{re}^{b,u,k}\right)\|_F}\right\|_2^2}{\sigma_n^2}\right).
\end{aligned}
\end{equation}
% \begin{equation}
%     \begin{aligned}
        
%     \end{aligned}
% \end{equation}
Since different $N_{l}$ influence the dimensionality of the precoding matrix, we have subdivided MIMO transmission tasks according to various $N_{l}$. Finally, the \textit{representative channel} between the BS $b$ and the UE $u$ can be represented as $\mathbf{H}_{re,*}^{b,u} = \bigcup_{k=1}^{N_k}\mathbf{H}_{re,*}^{b,u,k}\in \mathbb{C}^{N_k\times N_{R}\times N_T}$.
\subsubsection{Dual BS MIMO Transmission with Power Allocation}
For a fixed time $t$ and two BSs $b_1,b_2$, we fix $N_{l}=1$. Our objective is to maximize the throughput of the joint transmission with power allocation:
\begin{equation}\label{task2 ini obj}
\begin{aligned}
&\max\limits_{\mathbf{W}^{b_1,u,t,k},\mathbf{W}^{b_2,u,t,k}}\\&\sum_{k=1}^{N_k}
\log_2\left(1+\frac{\|\mathbf{H}^{b_1,u,t,k}\mathbf{W}^{b_1,u,t,k}+\mathbf{H}^{b_2,u,t,k}\mathbf{W}^{b_2,u,t,k}\|_2^2}{\sigma_n^2}\right)
\\&\text{s.t.} \sum_{k=1}^{N_k}\left(\left\Vert\mathbf{W}^{b_1,u,t,k}\right\Vert_{F} + \left\Vert\mathbf{W}^{b_2,u,t,k}\right\Vert_{F}\right)=2N_k,
\\& \forall b_1, b_2 \in  \left\{1,2,\cdots,N_b\right\}, \forall u\in  \left\{1,2,\cdots,N_u\right\}, b_1 \neq b_2.
\end{aligned}
\end{equation}
\textbf{Theorem 1.} The above problem is non-convex, but it can be transformed into the following convex problem and solved by the CVX tool, which indicates that optimal precoding matrix and power allocation can be addressed independently.
\begin{equation}\label{task2 convert}
    \begin{array}{cl}
&\max \limits_{P^{b_1,u,t,k}, P^{^{b_2,u,t,k}}} 
\\& \sum_{k=1}^{N_k} \log _2\left(1+\frac{1}{\sigma_n^2} (\lambda^{b_1,b_2,u,t,k})^2\left(P^{b_1,u,t,k}+P^{b_2,u,t,k}\right)\right) \\
\\
&\text { s.t. }\left\{\begin{array}{l} \sum_{k=1}^{N_k} P^{b_1,u,t,k}+ P^{b_2,u,t,k} \leq 2N_k,\\
P^{b_1,u,t,k} \geq 0, P^{b_2,u,t,k} \geq 0,\\\forall b_1, b_2\in  \left\{1,2,\cdots,N_b\right\}, \forall u\in  \left\{1,2,\cdots,N_u\right\}
\\\forall k\in  \left\{1,2,\cdots,N_k\right\}, b_1 \neq b_2.
\end{array}\right.
\end{array}
\end{equation}
where $\lambda^{b_1,b_2,u,t,k}$ is the biggest singular element of $\mathbf{H}^{b_1,b_2,u,t,k} = \left[\mathbf{H}^{b_1,u,t,k}, \mathbf{H}^{b_2,u,t,k}\right]\in \mathbb{C}^{N_{R}\times2N_T}$. $P^{b_1,u,t,k}=\left\Vert\mathbf{W}^{b_1,u,t,k}\right\Vert_{F}$ and $P^{b_2,u,t,k}=\left\Vert\mathbf{W}^{b_2,u,t,k}\right\Vert_{F}$ are powers for BSs $b_1$ and $b_2$ respectively. After solving the above problem, we can obtain the optimal power $P_{*}^{b_1,u,t,k}$ and $P_{*}^{b_2,u,t,k}$. Then, the optimal precoding matrix is given by
\begin{equation}\label{task2 solve}
\begin{aligned}
\mathbf{W}_{*}^{b_1,b_2,u,t,k}&=\mathscr{V}_1\left(\mathbf{H}^{b_1,b_2,u,t,k}\right)\hat{\Sigma}_{*}^{b_1,b_2,u,t,k}\\&=\left[\mathbf{W}_{*}^{b_1,u,t,k},\mathbf{W}_{*}^{b_2,u,t,k}\right]^T \in \mathbb{C}^{2N_T\times1},
\end{aligned}
\end{equation}
where $\hat{\Sigma}_{*}^{b_1,b_2,u,t,k}$ is the power allocation factor and it can be expressed as
\begin{equation}\label{power allocation result}
    \hat{\Sigma}_{*}^{b_1,b_2,u,t,k}=\left[\sqrt{P_{*}^{b_1,u,t,k}+ P_{*}^{{b_2,u,t,k}}},0,\ldots,0\right]^T \in \mathrm{C}^{2 N_t \times 1}.
\end{equation}
\textbf{Proof.} See Appendix A.

We denote $ \mathscr{U}_1\left(\mathbf{H}^{b_1,b_2,u,t,k}\right) \coloneqq \mathbf{W}_{*}^{b_1,u,t,k}$, and $\mathscr{U}_2\left(\mathbf{H}^{b_1,b_2,u,t,k}\right)\coloneqq \mathbf{W}_{*}^{b_2,u,t,k}$. Similarly, when we identify the \textit{representative channel} $\mathbf{H}_{re}^{b_1,b_2,u,k} \in \mathbb{C}^{N_{R}\times N_T}$, the problem is reformulated as 
\begin{equation}\label{task2 obj}
\begin{aligned}
&\mathbf{H}_{re,*}^{b_1,b_2,u,t,k}=\arg\max\limits_{\mathbf{H}_{re}^{b_1,b_2,u,t,k}}\text{TASK2}
\\&\operatorname{s.t.}\text{TASK2}=\\ &\frac{1}{N_kN_t}\sum_{t=1}^{N_t}\sum_{k=1}^{N_k}\log_2\left(1+\frac{\|\mathbf{H}^{b_1,u,t,k}\mathscr{U}_1\left(\mathbf{H}_{re}^{b_1,b_2,u,k}\right)}{}\right.\\&~~~~~~~~~~~~~~~~~~~~~~~~~~~~~\left.\frac{+\mathbf{H}^{b_2,u,t,k}\mathscr{U}_2\left(\mathbf{H}_{re}^{b_1,b_2,u,k}\right)\|_2^2}{\sigma_n^2}\right),
\end{aligned}
\end{equation}
% \begin{equation}
%     \begin{aligned}
        
%     \end{aligned}
% \end{equation}
and the \textit{representative channel} between BSs $b_1,b_2$ and UE $u$ can be represented as $\mathbf{H}_{re,*}^{b_1,b_2,u} = \bigcup_{k=1}^{N_k}\mathbf{H}_{re,*}^{b_1,b_2,u,k}\in \mathbb{C}^{N_k\times N_{R}\times 2N_T}$.
\section{Proposed Solution}
\begin{figure}[htbp]
\centerline{\includegraphics[width=0.87\linewidth]{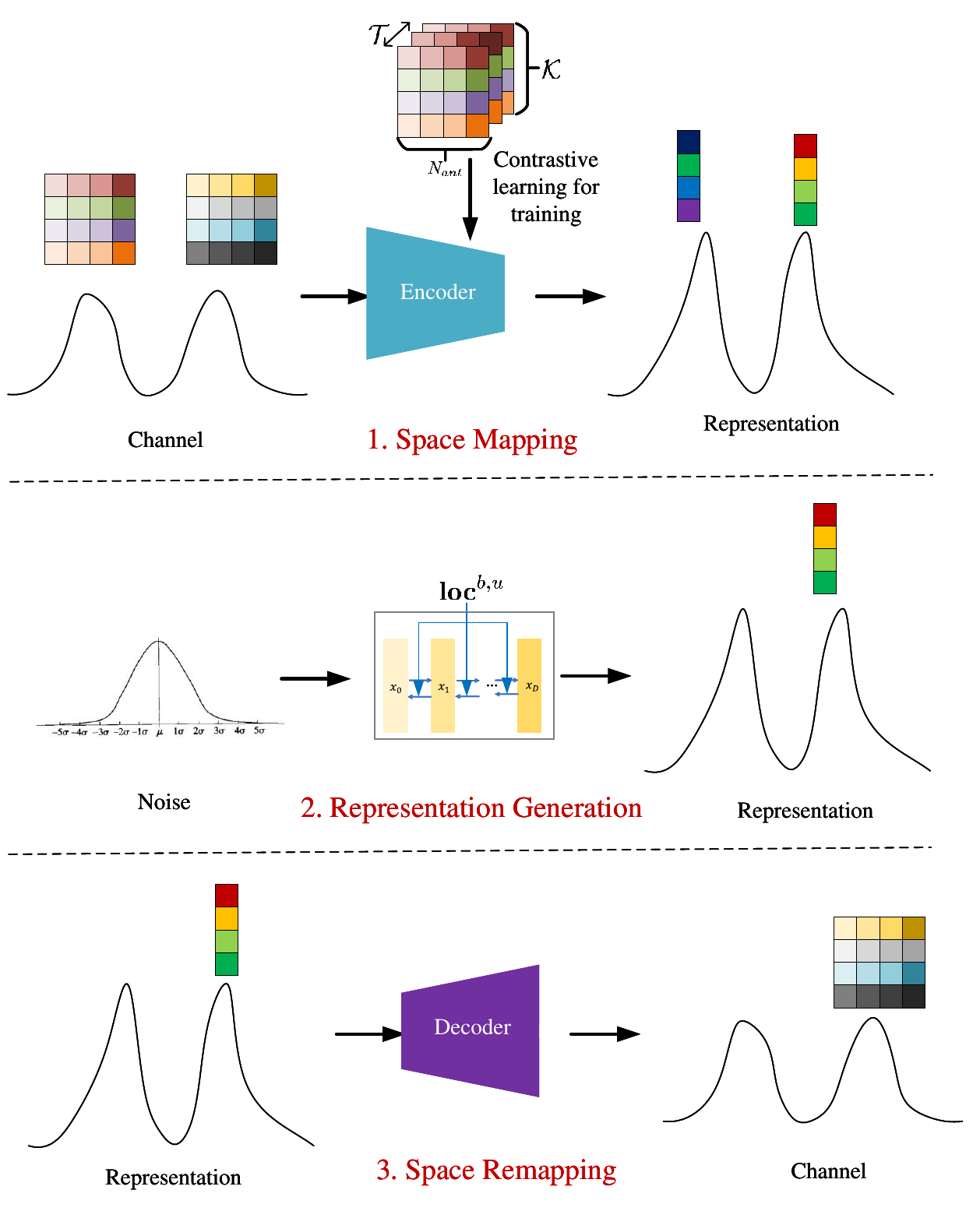}}
\caption{The overall framework for channel representation and generation.}
\label{fig:framework}
\end{figure}
In this section, we first introduce the overall framework of our work, which comprises three key components: an encoder obtained through contrastive learning, a conditional generator, and a decoder, as illustrated in Fig. \ref{fig:framework}. The encoder maps the channel matrix to the channel representation in latent space using self-supervised learning. Subsequently, a conditional generator, trained in the latent space, samples from the noise distribution to generate a representation conforming to the desired distribution. Finally, the decoder remaps the channel representation back to the channel matrices, suitable for various channel-related tasks.
\subsection{Encoder}
\begin{figure}[htbp]
\centerline{\includegraphics[width=0.9\linewidth]{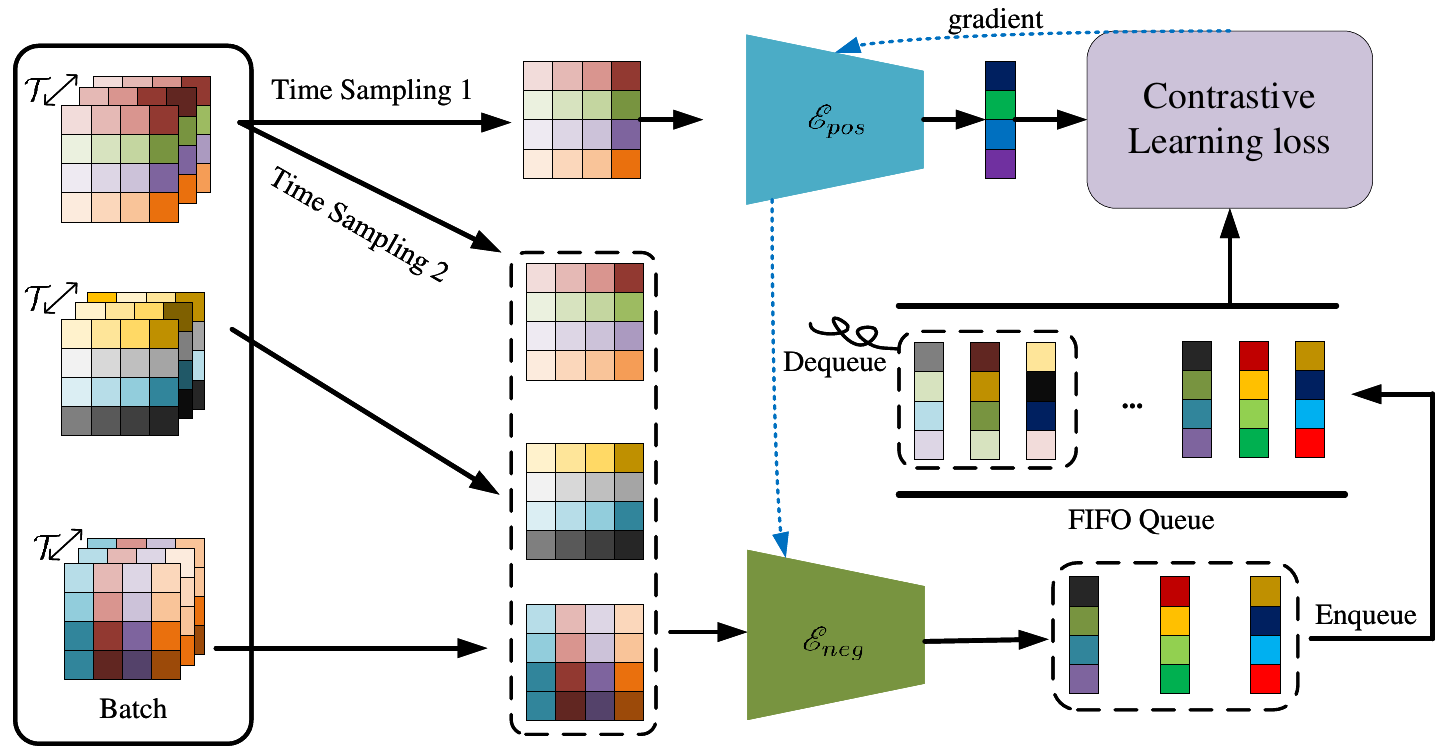}}
\caption{The training process of the encoder.}
\label{fig:encoder}
\end{figure}
\begin{figure}[htbp]
\centerline{\includegraphics[width=0.8\linewidth]{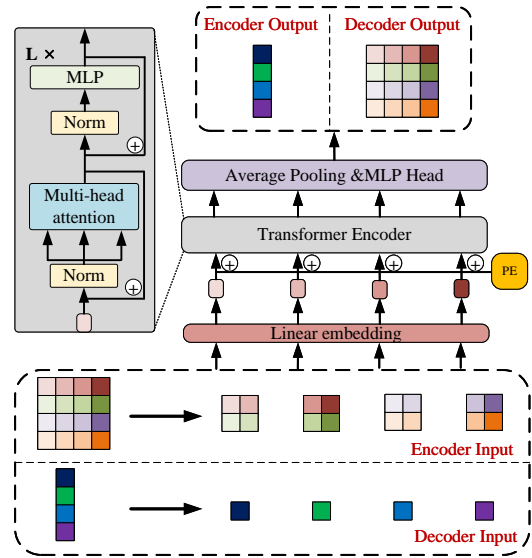}}
\caption{The structure of encoder and decoder based on Transformer encoder.}
\label{fig:decoder}
\end{figure}

The encoder is trained using contrastive learning as shown in Fig. \ref{fig:encoder}, where the fundamental idea is that channels between the same BS-UE pair are approximately similar, while channels between different BS-UE pair are dissimilar. We use $\{b,u\}_q$ to denote a BS-UE pair. Suppose there is one positive sample $\mathbf{H}^{\{b,u\}_0}$ and $\mathcal{Q}$ negative samples $\{\mathbf{H}^{\{b,u\}_q}\}_{\mathcal{Q}}(q\in\{1,2,\ldots,\mathcal{Q}\})$ in a training session, where 
$\mathbf{H}^{\{b,u\}_0, t}$ and $\mathbf{H}^{\{b,u\}_0, t'},\left(t,t'\in \{1,2,\ldots,N_t\}, t \neq t'\right)$ are approximately similar, while 
$\mathbf{H}^{\{b,u\}_0,t}, \mathbf{H}^{\{b,u\}_q,t'}$ are dissimilar. Therefore, we need to train two encoders, namely $\mathscr{E}_{pos}$ and $\mathscr{E}_{neg}$. $\mathscr{E}_{pos}\left(\mathbf{H}^{\{b,u\}_0, t}\right)$ and $\mathscr{E}_{neg}\left(\mathbf{H}^{\{b,u\}_0, t'}\right)$ are designed  to be as close as possible in the latent space, while $\mathscr{E}_{pos}\left(\mathbf{H}^{\{b,u\}_0, t}\right)$ and $\mathscr{E}_{neg}\left(\mathbf{H}^{\{b,u\}_q, t'}\right)$  are as far apart as possible. The loss function can be expressed as
\begin{equation}\label{eq: encoder loss}
\begin{aligned}
&\mathcal{L}_{\text{En}} \coloneqq \\&-\log_2\frac{\exp\left(\mathscr{E}_{pos}\left(\mathbf{H}^{\{b,u\}_0, t}\right)\cdot\mathscr{E}_{neg}\left(\mathbf{H}^{\{b,u\}_0, t'}\right)/\gamma\right)}{\sum_{q=0}^{\mathcal{Q}}\exp\left(\mathscr{E}_{pos}\left(\mathbf{H}^{\{b,u\}_0, t}\right)\cdot\mathscr{E}_{neg}\left(\mathbf{H}^{\{b,u\}_q, t'}\right)/\gamma\right)},
\end{aligned}
\end{equation}
where $\gamma$ is a temperature parameter that influences whether the representation distribution is peaked or smooth. 
The encoder is trained following the approach described in \cite{he2020momentum}. Instead of using the channel captured by each batch directly as a negative sample, we employ a first-in-first-out (FIFO) queue for negative samples. The output of $\mathscr{E}_{neg}$
  for each batch is enqueued into the FIFO queue, while the output of the earliest entry is dequeued from the FIFO queue. The parameters of $\mathscr{E}_{pos}$ are updated through standard backpropagation, while the parameters of $\mathscr{E}_{neg}$
  are gradually updated based on $\mathscr{E}_{pos}$'s parameters to ensure the consistency of the queue. The updating process of $\theta_{neg}$ can be expressed as
\begin{equation}\label{eq:encoder update}
    \theta_{neg} \leftarrow \beta\theta_{neg} + (1-\beta)\theta_{pos},
\end{equation}
where $\beta \in [0,1)$ and it is typically set to a value very close to 1. It is highlighted that CNNs are not suitable for channel matrices due to the lack of translation invariance\cite{chen2023viewing}. Therefore, we employ a Transformer encoder structure, decompose the channel into patches in the antenna and frequency domains, and embed them into $d_{model}$ dimensions  as shown in Fig. \ref{fig:decoder}. Subsequently, sine/cosine positional encoding is applied:
\begin{equation}\label{PE}
\begin{aligned}
&\text{PE}(p o s, 2 i)  =\sin \left(p o s / 10000^{2 i / d_{\text {model }}}\right), \\
&\text{PE}(p o s, 2 i+1)  =\cos \left(p o s / 10000^{2 i / d_{\text {model }}}\right),
\\&\mathscr{P}(pos)=[\text{PE}(p o s, 0),\text{PE}(p o s, 1),\ldots,\text{PE}(p o s, d_{model}-1)],
\end{aligned}    
\end{equation}
where $pos$ is the positional index and $i$ is the demension index. PE and $\mathscr{P}$ are the output of the positional encoding. The input features are then extracted by a multilayer Transformer encoder with a self-attention module, which maps the input to a dimension of size $d_{atten}$ through three independent linear layers with weights $\mathbf{\Omega}_\Theta$, $\mathbf{\Omega}_K$ and $\mathbf{\Omega}_V$ respectively. Then the output, denoted by $\mathbf{\Theta}, \mathbf{K}$ and $\mathbf{V}$, performs the attention computation as follows:
  \begin{equation}\label{eq:self-attention}
  \begin{split}
&\mathbf{\Theta}=\mathbf{\Omega}_\Theta\cdot \text{input},\mathbf{K}=\mathbf{\Omega}_K\cdot \text{input},\mathbf{V}=\mathbf{\Omega}_V\cdot \text{input},
\\&\mathscr{A}(\mathbf{\Theta},\mathbf{K},\mathbf{V}) = \operatorname{softmax}\left(\frac{\mathbf{\Theta}\mathbf{K}^T}{\sqrt{d_{atten}}}\right)\mathbf{V},
\end{split}
  \end{equation}
where $\mathscr{A}$ is attention computation and $\operatorname{softmax}$ is the softmax operation\cite{goodfellow2016deep}. The transformer-encoder structure is repeated $L$ times and the final feature output is obtained through average pooling and an multilayer perceptron (MLP) head. Additionally, an MLP projector is appended to the encoder during training to enhance the performance of contrastive learning\cite{chen2020simple}. The detailed procedures of the proposed encoder is given in Algorithm~1.

\begin{algorithm}\label{alg:algorithm1}
    \caption{Encoder algorithm.}\label{algorithm1}
$\textbf{Input:}$ Dataset $\left\{\mathbf{H}^{b,u}\right\}_{ N_b\times N_u}$.\\
Initializes $\theta_{pos}$ and $\theta_{neg} \leftarrow \theta_{pos}$ \\
\Repeat{converged}
    {
    Sample $\mathbf{H}^{b,u}$ from $\left\{\mathbf{H}^{b,u}\right\}_{ N_b\times N_u}$.\\
    Sample $\mathbf{H}^{b,u,t}$ and $\mathbf{H}^{b,u,t'}$ from $\mathbf{H}^{b,u}$\\
    Calculate gradient based on Eq. (\ref{eq: encoder loss}).\\
    Take gradient descent step for $\theta_{pos}$ based on the gradient.\\
    Update $\theta_{neg}$ based on Eq. (\ref{eq:encoder update}).\\
    Enqueue $\mathscr{E}_{neg}\left(\mathbf{H}^{\{b,u\}_0, t'}\right)$ into the FIFO.\\
    Dequeue the FIFO.
    }
\For{$\mathbf{H}^{b,u,t}$ in $\left\{\mathbf{H}^{b,u}\right\}_{ N_b\times N_u}$}
    {
    $\mathbf{F}^{b,u,t}= \mathscr{E}_{pos}(\mathbf{H}^{b,u,t})$
    }
$\textbf{Utilization:}$ Generate channel representation for the dataset $\left\{\mathbf{H}^{b,u}\right\}_{ N_b\times N_u}$
\end{algorithm}
\subsection{Generator}
\begin{figure}[htbp]
\centerline{\includegraphics[width=0.75\linewidth]{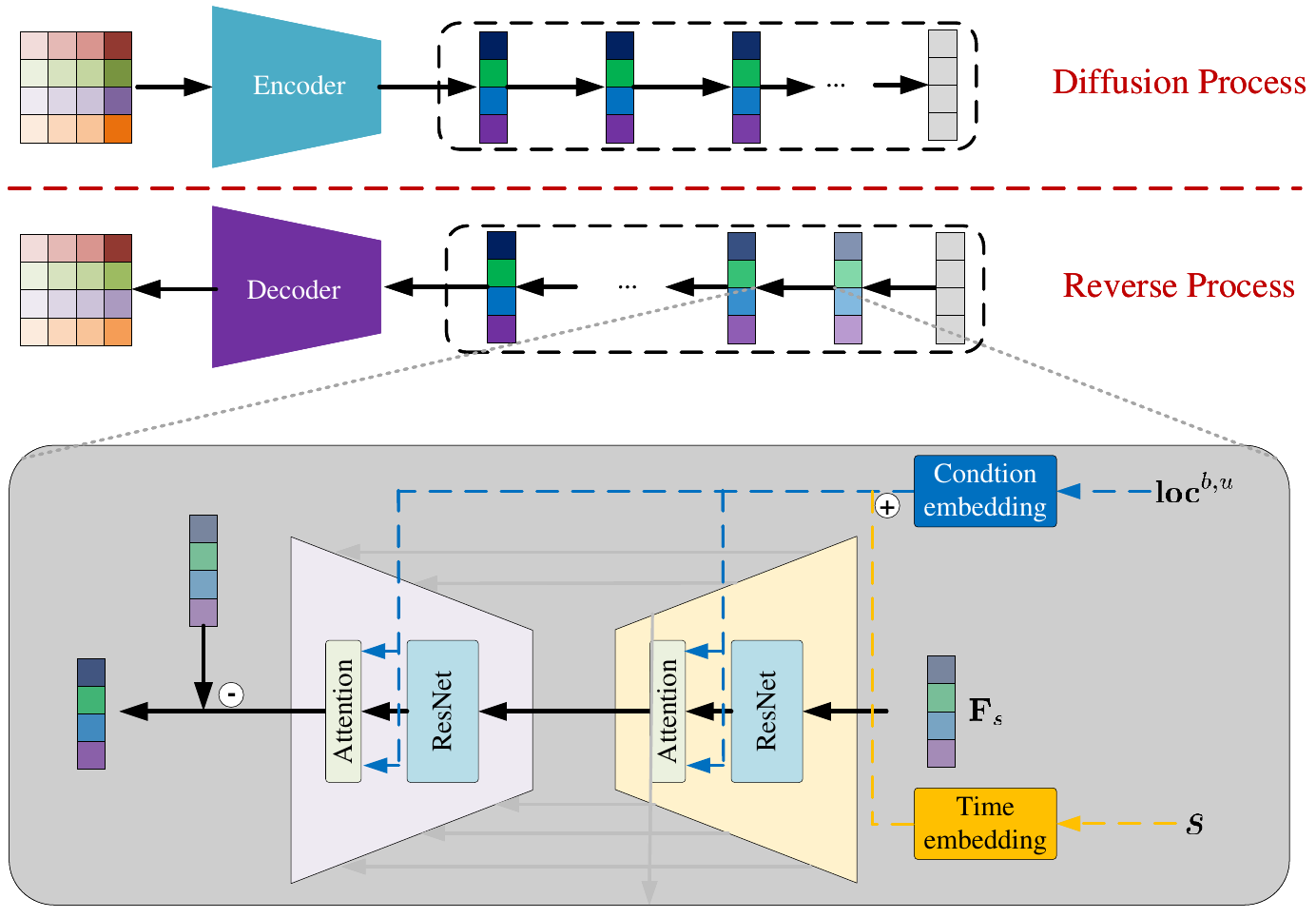}}
\caption{The overall structure of conditional diffusion model.}
\label{fig:diffusion}
\end{figure}
We employ geolocation information as a condition and train a conditional diffusion model as the generator. For location $\mathbf{loc}^{b,u}$
, we define $\mathbf{F}_0 \coloneqq \mathbf{F}^{b,u,t} \sim f(\mathbf{F}|\mathbf{loc}^{b,u})$ as the input of diffusion process. In the diffusion process, we add Gaussian noise with variance $\delta_s$ to this distribution for each step $s$. This process can be modeled as a Markov chain, and the joint distribution after $N_S$ steps can be represented as
\begin{equation}\label{joint distribution}
\begin{aligned}
f\left(\mathbf{F}_{1:N_S}|\mathbf{F}_{0},\mathbf{loc}^{b,u}\right)=&\prod_{s=1}^{N_S}f\left(\mathbf{F}_{s}|\mathbf{F}_{s-1},\mathbf{loc}^{b,u}\right)
\\=&\prod_{s=1}^{N_S}\mathcal{N}\left(\mathbf{F}_s;\sqrt{1-\delta_s}\mathbf{F}_{s-1},\delta_s\mathbf{I}\right).
\end{aligned}
\end{equation}
\cite{Ho2020Denoising} proved that $\mathbf{F}_s$ can be calculated using $\mathbf{F}_0$ and $\delta_s$:
\begin{equation}\label{eq:diffusion forward}
\begin{split}
% \mathbf{F}_s = \sqrt{\bar{\alpha}_{s}}\mathbf{F}_0 + \sqrt{1-\bar{\alpha}_{s}}\mathbf{z}_s
% \\
\mathbf{F}_s\sim& f\left(\mathbf{F}_s|\mathbf{F}_{0},\mathbf{loc}^{b,u}\right)\\=&f\left(\mathbf{F}_s|\mathbf{F}_{0}\right)
\\=&\mathcal{N}\left(\mathbf{F}_s;\sqrt{\bar{\alpha}_{s}}\mathbf{F}_0,1-\bar{\alpha}_{s}\mathbf{I}\right),
\end{split}
\end{equation}
where $\alpha_s\coloneqq1-\delta_s, \bar{\alpha}\coloneqq\prod_{s'=0}^{s}\alpha_{s'}$.

During the reverse diffusion process, we construct a parameterized Gaussian distribution to estimate the true distribution. The reverse diffusion process $p_{\omega}\left(\mathbf{F}_{s-1}|\mathbf{F}_{s}, \mathbf{loc}^{b,u}\right)$ can also be modeled as a Markov chain as follows:
\begin{equation}\label{eq: reverse process}
p_{\omega}\left(\mathbf{F}_{s-1}|\mathbf{F}_{s}, \mathbf{loc}^{b,u}\right)\!=\!\mathcal{N}\left(\mathbf{F}_{s-1};\boldsymbol{\mu}_{\omega}\left(\mathbf{F}_s,s,\mathbf{loc}^{b,u}\right),b_s\mathbf{I}\right)_{,}
\end{equation}
\begin{equation}\label{eq: reverse joint distribution}
    p_{\omega}\left(\mathbf{F}_{0:N_S}\right)=p_{\omega}\left(\mathbf{F}_{N_S}\right)\prod_{s=1}^{N_S}p_{\omega}\left(\mathbf{F}_{s-1}|\mathbf{F}_{s}\right),
\end{equation}
where $\boldsymbol{\mu}_{\omega}$ is the mean of Gaussian distribution with parameters $\omega$. $b_s$ is the variance, which depends on step 
$s$. In \cite{Ho2020Denoising}, it has been proved that 
$p_{\omega}\left(\mathbf{F}_{s-1}|\mathbf{F}_{s},\mathbf{F}_0\right)$ can be expressed by $\mathbf{F}_s$ and $\mathbf{F}_0$ with mean $\tilde{\boldsymbol{\mu}}_s$ and variance $\widetilde{\delta}_s$:
\begin{equation}\label{eq:DDPM thorem2}
    p_{\omega}\left(\mathbf{F}_{s-1}|\mathbf{F}_{s},\mathbf{F}_0\right)=\mathcal{N}\left(\mathbf{F}_{s-1};\widetilde{\boldsymbol{\mu}}_s(\mathbf{F}_{s},\mathbf{F}_0),\widetilde{\delta}_s\mathbf{I}\right),
\end{equation}
where
\begin{equation}\label{eq:DDPM th3}\begin{aligned}\tilde{\boldsymbol{\mu}}_s\left(\mathbf{F}_s, \mathbf{F}_0\right) & =\left(\frac{\sqrt{\alpha_s}}{\delta_s} \mathbf{x}_s+\frac{\sqrt{\bar{\alpha}_s}}{1-\bar{\alpha}_s} \mathbf{F}_0\right) /\left(\frac{\alpha_s}{\delta_s}+\frac{1}{1-\bar{\alpha}_{s-1}}\right)\\&=\frac{\sqrt{\alpha_s}\left(1-\bar{\alpha}_{s-1}\right)}{1-\bar{\alpha}_s} \mathbf{F}_s+\frac{\sqrt{\bar{\alpha}_{s-1}} \delta_s}{1-\bar{\alpha}_s} \mathbf{F}_0
\\&=\frac{1}{\sqrt{\alpha_s}}\left(\mathbf{F}_s-\frac{\delta_s}{\sqrt{1-\bar{\alpha}_s}}\mathbf{z}_s\right),
\end{aligned}
\end{equation}
\begin{equation}\label{eq: DDPM variance}
\widetilde{\delta}_s =1 /\left(\frac{\alpha_s}{\delta_s}+\frac{1}{1-\bar{\alpha}_{s-1}}\right)=\frac{1-\bar{\alpha}_{s-1}}{1-\bar{\alpha}_s} \cdot \delta_s,
\end{equation}
where $\mathbf{z}_s \sim \mathcal{N}\left(\mathbf{0}, \mathbf{I}\right)$ is a standard multivariate Gaussian distribution. We use a noise prediction network $\mathscr{N}_{\omega}$ to predict $\mathbf{z}_s$:
\begin{equation}\label{eq:construct Unet}
\begin{aligned}
&\boldsymbol{\mu}_{\omega}\left(\mathbf{F}_s,s,\mathbf{loc}^{b,u}\right)=\\&~~~~~~~~~\frac{1}{\sqrt{\alpha_s}}\left(\mathbf{F}_s-\frac{\delta_s}{\sqrt{1-\bar{\alpha}_s}}\mathscr{N}_{\omega}\left(\mathbf{F}_s,s,\mathbf{loc}^{b,u}\right)\right).
\end{aligned}
\end{equation}
The optimization for the noise prediction network $\mathscr{N}_{\omega}$ involves the variational bound on the negative log-likelihood of $p_\omega\left(\mathbf{F}_0 \mid \mathbf{loc}^{b,u}\right)$ as the loss function:
\begin{equation}\label{eq: loss gen}
\begin{aligned}
&\mathbb{E}\left[-\log p_\omega\left(\mathbf{F}_0 \mid \mathbf{loc}^{b,u}\right)\right] \leq \\&\mathbb{E}_f\left[-\log \frac{p_\omega\left(\mathbf{F}_{0: N_S} \mid \mathbf{loc}^{b,u}\right)}{f\left(\mathbf{F}_{1: N_S} \mid \mathbf{F}_0, \mathbf{loc}^{b,u}\right)}\right]=: \mathcal{L}\left(\mathscr{N}_\theta\right).
\end{aligned}
\end{equation}
It was demonstrated in \cite{Ho2020Denoising} that the loss function can be simplified to the following form:
\begin{equation}\label{eq: loss gen simple}
\begin{aligned}
\mathcal{L}_{\text{Gen}} \coloneqq& \mathbb{E}_{s, \mathbf{loc}^{b,u}, \mathbf{F}_0, \mathscr{N}_s}\\
&\left[\left\|\mathscr{N}_s-\mathscr{N}_\omega\left(\sqrt{\bar{\alpha}_s} \mathbf{F}_0+\sqrt{1-\bar{\alpha}_s} \mathscr{N}_s, s, \mathbf{loc}^{b,u}\right)\right\|^2\right],
\end{aligned}    
\end{equation}
where $\mathscr{N}_s \sim \mathbf{z}_{s}=\mathcal{N}(\boldsymbol{0},\mathbf{I})$. We employ a U-Net structure to construct this network \cite{ronneberger2015u}, and we add a cross-attention layer following each convolutional layer to incorporate the condition as shown in Fig. \ref{fig:diffusion}. The method for time embedding is consistent with the positional encoding in the encoder, using $\mathscr{P}(s)$ for the embedding process. We denote the output of the CNN as $\mathbf{F}'_s$. Then, the output $\mathbf{\Theta}', \mathbf{K}'$, and $\mathbf{V}'$ for cross attention layer with weights $\mathbf{\Omega}'_\Theta, \mathbf{\Omega}'_K$ and $\mathbf{\Omega}'_V$ can be expressed as
\begin{equation}\label{eq:cross attention}
\begin{split}
&\mathbf{\Theta}'=\mathbf{\Omega}'_\Theta\cdot \mathbf{F}'_s,
\\&\mathbf{K}'=\mathbf{\Omega}'_K\cdot (\mathscr{P}(s)+\mathscr{C}(\mathbf{loc}^{b,u})),
\\&\mathbf{V}'=\mathbf{\Omega}'_V\cdot  (\mathscr{P}(s)+\mathscr{C}(\mathbf{loc}^{b,u})),
\end{split}
\end{equation}
where $\mathscr{C}(\mathbf{loc}^{b,u})$ is the embedding vector for condition $\mathbf{loc}^{b,u}$. The detailed procedures of the proposed generator are given in Algorithm 3.

\begin{algorithm}\label{alg:diffusion}
    \caption{Generator algorithm.}
$\textbf{Input:}$ Dataset $\left\{\mathbf{loc}^{b,u},\mathbf{F}^{b,u}\right\}_{ N_b\times N_u}$.\\
\Repeat{converged}
    {
    Sample $\mathbf{loc}^{b,u}$ and $\mathbf{F}^{b,u}$ from $\left\{\mathbf{loc}^{b,u},\mathbf{F}^{b,u}\right\}_{ N_b\times N_u}$.\\
    Sample an original channel $\mathbf{F}^{b,u,t}$ from $\mathbf{F}^{b,u}$.\\
    Sample $s \sim \text{Uniform}(\{1,\ldots,N_S\})$.\\
    Sample $\mathscr{N}_s \sim \mathcal{N}(\boldsymbol{0},\mathbf{I})$.\\
    Take gradient descent step based on Eq. (\ref{eq: loss gen simple})
    
    }
Sample $\mathbf{F}_{N_S}\sim \mathcal{N}(\boldsymbol{0},\mathbf{I})$\\
\For{$s = N_S,\ldots,1$}
    {
    Sample $\mathbf{z}_s \sim \mathcal{N}\left(\mathbf{0}, \mathbf{I}\right)$ if $s > 1$, else $\mathbf{z}_s = 0$.\\
    Obtain $\mathbf{F}_{s-1}$ based on Eqs. (\ref{eq:DDPM thorem2}), (\ref{eq:DDPM th3}) and (\ref{eq: DDPM variance})
    }
$\textbf{Utilization:}$ Generate 
 channel representation based on location information.
\end{algorithm}

 \subsection{Decoder}
To ensure that the representation can be accurately recovered to the channel matrix, the decoder employs the same Transformer-encoder structure as the encoder. However, the patches in the decoder are divided into smaller segments, and the dimensions are gradually expanded through the MLP layer to ultimately recover the channel matrix. In the CNN-based structure, the encoder performs convolution, and the decoder typically employs a deconvolution operation. However, this can lead to the checkerboard effect, which is undesirable for the channel matrix. Instead, we utilize a Transformer-encoder structure,  which is better suited for handling channels compared to a CNN structure as shown in Fig. \ref{fig:decoder}. After the training process of encoder, it can obtain the representation of the entire dataset, as defined in $\mathbf{F}^{b,u,t}\coloneqq \mathscr{E}_{pos}(\mathbf{H}^{b,u,t}) \in \mathbb{C}^{N_{re}\times 1}$, where $N_{re}$ is the number of dimensions of representation. Using this representation, the dataset 
$\{\mathbf{F}^{b,u,t},\mathbf{H}^{b,u,t}\}_{ N_b\times N_u\times{N_t}}$ can be obtained and utilized for training the decoder. The goal of the decoder is to remap the representation back to the channel. Therefore, it is trained using the MSE loss:
\begin{equation}\label{eq:decoder loss}
    \mathcal{L}_{\text{de}} \coloneqq \sum_{b=1}^{N_b}\sum_{u=1}^{N_u}\sum_{t=1}^{T}\left\|\mathbf{H}^{b,u,t} - \mathscr{D}\left(\mathbf{F}^{b,u,t}\right)\right\|^2,
\end{equation}
where $\mathscr{D}$ presents the Transformer-encoder-based decoder network. The detailed procedures of the proposed decoder is given in Algorithm~2.
\begin{algorithm}\label{alg:decoder}
    \caption{Decoder algorithm.}
$\textbf{Input:}$ Dataset $\{\mathbf{F}^{b,u,t},\mathbf{H}^{b,u,t}\}_{ N_b\times N_u\times{N_t}}$.\\
\Repeat{converged}
    {
    Sample $\mathbf{F}^{b,u,t}$ and $\mathbf{H}^{b,u,t}$ from $\{\mathbf{F}^{b,u,t},\mathbf{H}^{b,u,t}\}_{ N_b\times N_u\times{N_t}}$.\\
    Take gradient descent step based on Eq. (\ref{eq:decoder loss})
    
    }
$\textbf{Utilization:}$ Remap representation $\mathbf{F}^{b,u,t}$ to channel $\mathbf{H}^{b,u,t}$.
\end{algorithm}
\section{Simulation Results}

In this section, we introduce a vector quantized variational autoencoder (VQ-VAE)-based representation and generation method as a baseline. Next, we outline our simulation parameter settings and conduct ablation experiments to assess the performance of the Transformer-encoder based model on the channel. Finally, we validate the simulation results through visualization and geolocation-based MIMO transmission for both representation and generation tasks.
\subsection{Baselines}
First, a straightforward baseline involves traversing the $N_t$ original channels in the dataset and selecting the one that achieves the highest scores in $\text{TASK1}_{N_{l}}$ and $\text{TASK2}$ as the representative channel. 

\begin{figure}[htbp]
\centerline{\includegraphics[width=0.9\linewidth]{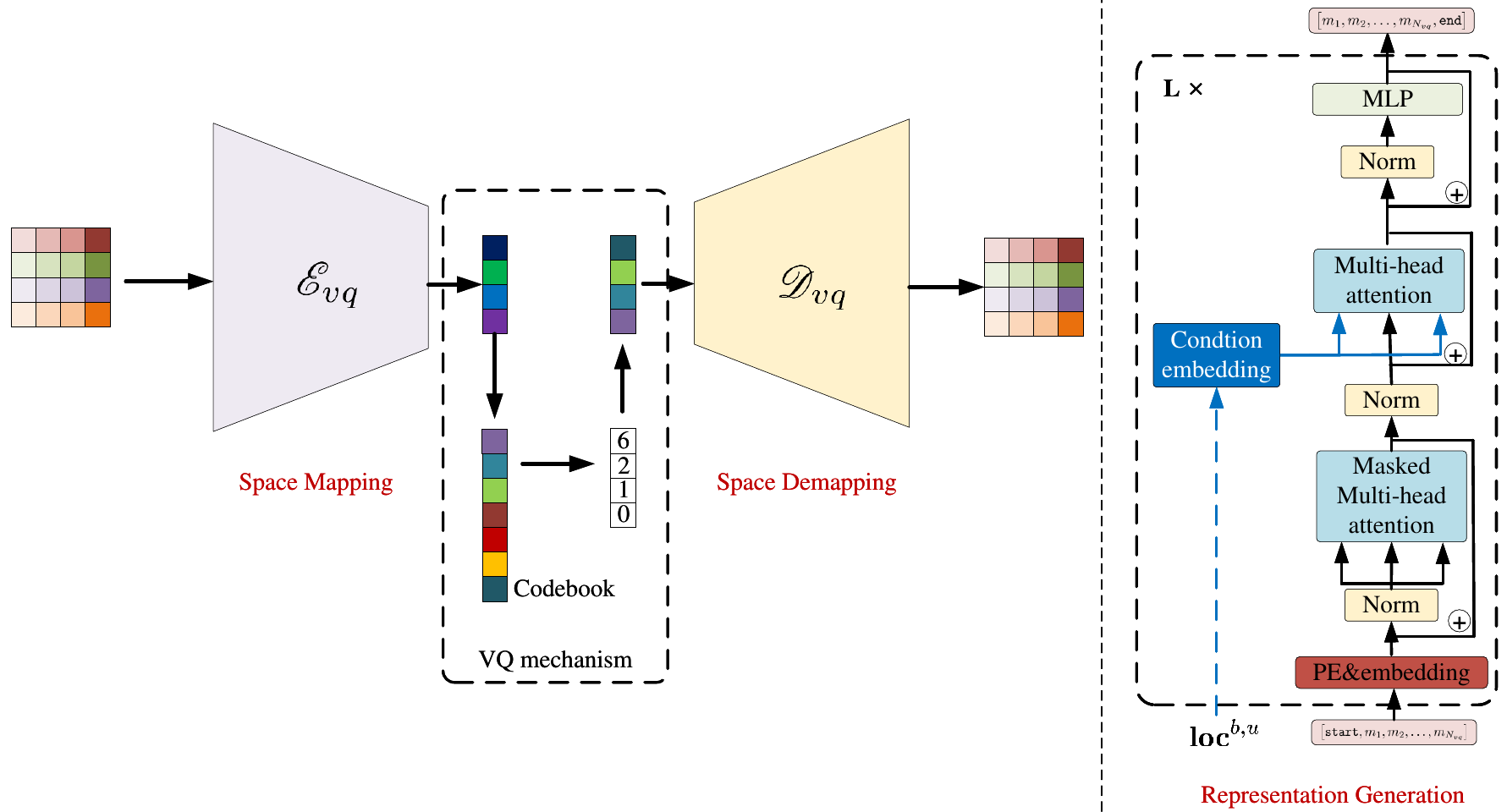}}
\caption{The baseline method for channel representation and generation that includes a VQ-VAE and a GPT-like autogressive model.}
\label{fig:baseline}
\end{figure}

Additionally, we employ a VQ-VAE-based approach as a performance baseline. Note that this approach has demonstrated significant success in CV\cite{van2017neural, ramesh2021zero}, as well as the compression-recovery task of channel\cite{shin2024vector}. VQ-VAE trains both an encoder and a decoder while maintaining a codebook, thereby enabling the mapping of input data to a discrete latent space\cite{van2016conditional}. Once the encoder and decoder are trained, we utilize an autoregressive network to predict the index of the codebook and realize generation given that the representation is discrete. Unlike the original VQ-VAE paper, which utilized PixelCNN for generation, We employ a pre-trained transformer-like (GPT-like) network that adopts only the Transformer decoder structure\cite{brown2020language,radford2019language}, augmented with a cross-attention mechanism to incorporate the condition. The cross-attention mechanism utilizes the method described in Eq.~\eqref{eq:cross attention}. Additionally, \texttt{start} and \texttt{end} identifiers are added for training the autoregressive network\cite{vaswani2017attention}, with the specific network structure detailed in Fig. \ref{fig:baseline}.

\subsection{Simulation Settings}
\begin{figure}[ht]
  \centering
  \subfloat[The DeepMIMO $\texttt{O1\_3p5}$ scenario.]
  {
    \includegraphics[width=0.45\linewidth]{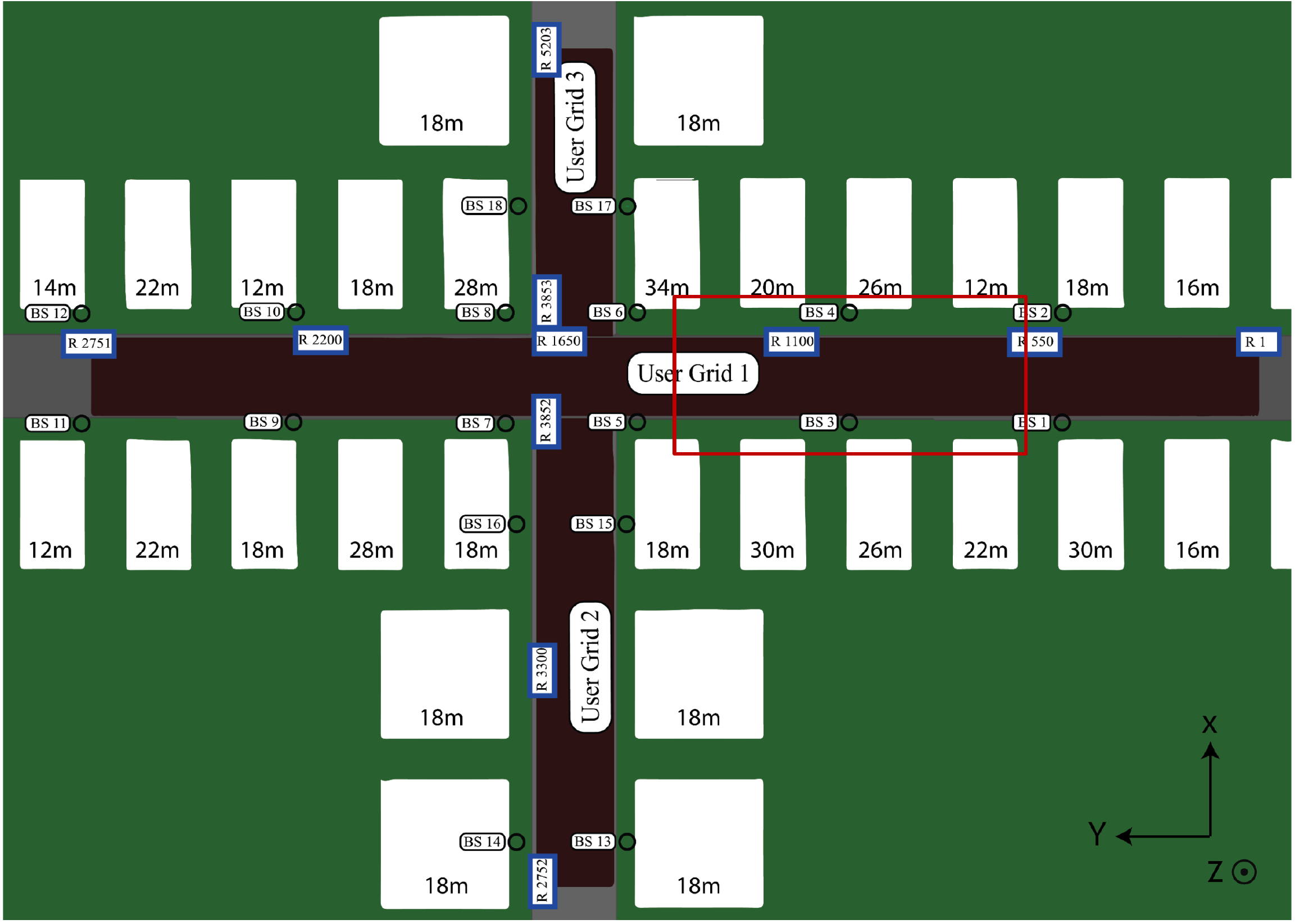}
    \label{fig:1}
  }
  \subfloat[The geolocations of UEs and BSs.]
  {
    \includegraphics[width=0.452\linewidth]{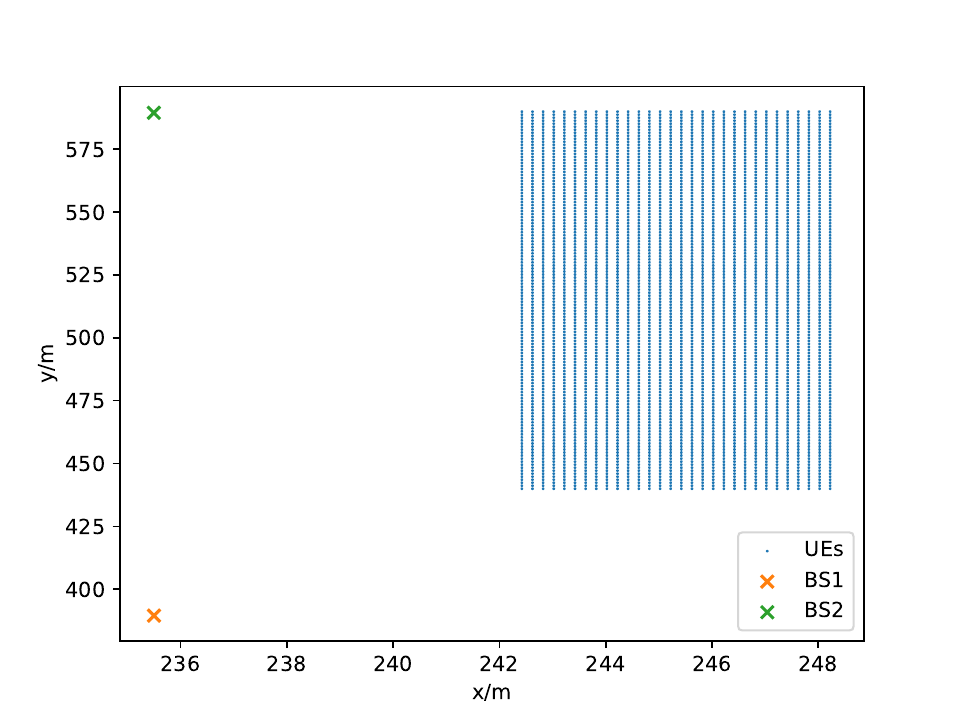}
    \label{fig:2}
  }
  \caption{The selected scenario of DeepMIMO dataset and chosen locations of UEs and BSs.}
  \label{fig:deepmimo}
\end{figure}

% \begin{figure}[htbp]
% \centerline{\includegraphics[width=0.8\linewidth]{figures/Deep MIMO_O1.pdf}}
% \caption{The DeepMIMO $\texttt{O1\_3p5}$ scenario.}
% \label{fig:DeepMIMO O1}
% \end{figure}

\begin{table}[htbp]
    \centering
    \caption{Simulation Parameters of DeepMIMO Dataset.}
    \begin{tabular}{ll}
        \toprule
        Parameter  & Value \\
        \midrule
        % Channel Scenario    & O1\_3p5\\
        Candidate BS Index  & 1, 5  \\
        Candidate UE Index  & 715-1465 rows\\  
        BS Antenna Configuration & 8$\times$2 ports with two polarization\\
        UE Antenna Configuration & 4 port\\
        % $ N_b$       &  1.92MHz\\
        % Number of Subcarriers    & 128\\
        % $N_t$    & 50\\
        \bottomrule
    \end{tabular}
    \label{tb:deepMIMO settings}
\end{table}

\begin{table}[htbp]
    \centering
    \caption{Hyperparameters of Encoder and Decoder.}
    \begin{tabular}{c|cc}
        \toprule
        Parameter  & Encoder value & Decoder value \\
        \midrule
        $\texttt{input\_size}$      & (2,128,128) & (2,8,8)\\
        $\texttt{patch\_size}$      & 16 & 1\\
        $d_{model}$      & 512 & 512\\
        number of multi-head      & 6 & 12\\
        $L$                       & 4 & 8\\
        \bottomrule
    \end{tabular}
    \label{tb:Hyperparameters en-de}
\end{table}

% \begin{minipage}{\linewidth}
% \begin{minipage}[t]{0.48\linewidth}
% \makeatletter\def\@captype{table}
% % \begin{table}[htbp]
% %     \centering
% %     \caption{Hyperparameters of MoCo Training}
    
%     \caption{Hyperparameters of MoCo Training}
%     \label{tb:Hyperparameters en-de}
% % \end{table}
% \end{minipage}
% \begin{minipage}[t]{0.48\linewidth}
% \makeatletter\def\@captype{table}
% % \begin{table}[htbp]
% %     \centering
% %     \caption{Hyperparameters of VQ-VAE based method}
%     \begin{tabular}{c|c}
%         \toprule
%         Parameter  & Value\\
%         \midrule
%         $N_{vq}$& 4\\
%         $N_\chi$      & 32\\
%         $N_c$      & 8192\\
%         \bottomrule
        
%     \end{tabular}
%     \caption{Hyperparameters of VQ-VAE based method}
%     \label{tb:Hyperparameters VQVAE}
% % \end{table}
% \end{minipage}
% \end{minipage}

% \begin{figure}[htbp]
% \centerline{\includegraphics[width=0.8\linewidth]{figures/dataset_loc.pdf}}
% \caption{The The geolocations of UEs and BSs.}
% \label{fig:DeepMIMO loc}
% \end{figure}
We evaluate the simulation performance of our proposed method using a channel dataset generated by real ray tracing. As illustrated in Fig. \ref{fig:deepmimo}, we obtain the ray tracing results in the \texttt{O1\_3p5} scenario of the DeepMIMO dataset, and the channels are generated in MATLAB based on the clustered delay line (CDL) model and Eq. \eqref{eq:channel model}. We configure the transmitter with 32 antenna ports and the receiver with 4 antenna ports, utilizing a total of 128 subcarriers and a bandwidth of 1.92 MHz. CDL channels are generated at 50 independent time instants between 3030 UEs and 2 BSs to capture variations in the temporal domain, resulting in a total of 303,000 channels. The additional setup parameters are detailed in Table \ref{tb:deepMIMO settings}. 
\subsection{Ablation of Transformer Encoder Structure}
As previously discussed, the inductive bias of convolution operations and the checkerboard effect from deconvolution operations are theorized to degrade performance when neural networks are applied to channel matrices. In this section, we conduct ablation studies to test this hypothesis and verify performance. The specific hyperparameters for our proposed encoder and decoder, which are based on the Transformer encoder structure, are detailed in Table~\ref{tb:Hyperparameters en-de}. Additionally, we implement an encoder based on ResNet-18 and a corresponding decoder based on a Deconvolutional Network\cite{he2016deep}. These two encoder-decoder pairs, while employing different structures, have approximately the same number of parameters, as calculated by the \texttt{torchsummary} module. Then, we evaluate their performance based on $\text{TASK1}_1$ for the compression-recovery task. Specifically, this task involves passing the channel through an encoder to obtain a representation, and then remapping this representation to the channel using a decoder. After recovering the $N_t$ channels, we traverse them and select the one that achieves the highest score in the task. Another metric to quantify the difference between the original channel and recovered one is NMSE, which can expressed as follows:
 \begin{equation}
 \begin{aligned}
 &\text{NMSE}=10\log_{10}\\
 &~~~~\left(\frac{1}{ N_b N_uN_t}\sum_{b=1}^{N_b}\sum_{u=1}^{ N_u}\sum_{t=1}^{N_t}\frac{\|\mathbf{H}^{b,u,t}-\mathscr{D}\left(\mathscr{E}(\mathbf{H}^{b,u,t})\right)\|_2^2}{\|\mathbf{H}^{b,u,t}\|_2^2}\right).
 \end{aligned}
 \end{equation}
 The results indicate that the encoder-decoder with the CNN structure does not fully converge, resulting in a significant discrepancy between the decoder's output and the encoder's input, while Transformer-encoder-based encoder-decoder achieves performance on $\text{TASK1}_1$ that is nearly identical to that of the original channel as shown in Table \ref{tb:ablation}.
\begin{table}[htbp]
	\caption{Ablation of CNN and Transformer Encoder Structure.}	\centering
	\begin{tabular}{p{3.4cm}|ccc}
		\toprule
Method & \texttt{\#param} & NMSE(-dB) &$\text{TASK1}_1$
\\
\midrule 
Encoder with CNN backbone & 85.75M &\multirow{2}{*}{-11.54} & \multirow{2}{*}{3.962}\\
Decoder with CNN backbone & 85.75M &&\\
\midrule 
Encoder with Transformer encoder backbone & 74.64M & \multirow{2}{*}{-20.13} & \multirow{2}{*}{4.933}\\
Decoder with Transformer encoder backbone & 74.04M & & \\
\midrule 
Orginal channel &- & -&4.936\\
		\bottomrule
	\end{tabular}
 \label{tb:ablation}
\end{table}

\subsection{Representation Performance}
\begin{table}[htbp]
    \centering
    
    \begin{tabular}{c|c}
        \toprule
        Parameter  & Value\\
        \midrule
        Length of FIFO queue & 16384\\
        $\beta$      & 0.99\\
        $\gamma$      & 5e-3\\
        $N_{re}$    & 128\\
        Learning rate & 5e-5 (with linear warmup)\\
        \midrule
        Number of vector & 4\\
        Number of dimensions of vector      & 32\\
        Size of codebook     & 8192\\
        Learning rate & 5e-6 (with linear warmup)\\
        \bottomrule
    
    \end{tabular}
    \caption{Hyperparameters of the Encoder for our Method and the VQ-VAE-based Method in the Training Process.}
    \label{tb:Hyperparameters encoder}
\end{table}
\begin{figure*}[htbp]
\begin{minipage}{0.74\textwidth}
  \centering
  \subfloat[Channels.]
  {
    \includegraphics[width=0.3\linewidth]{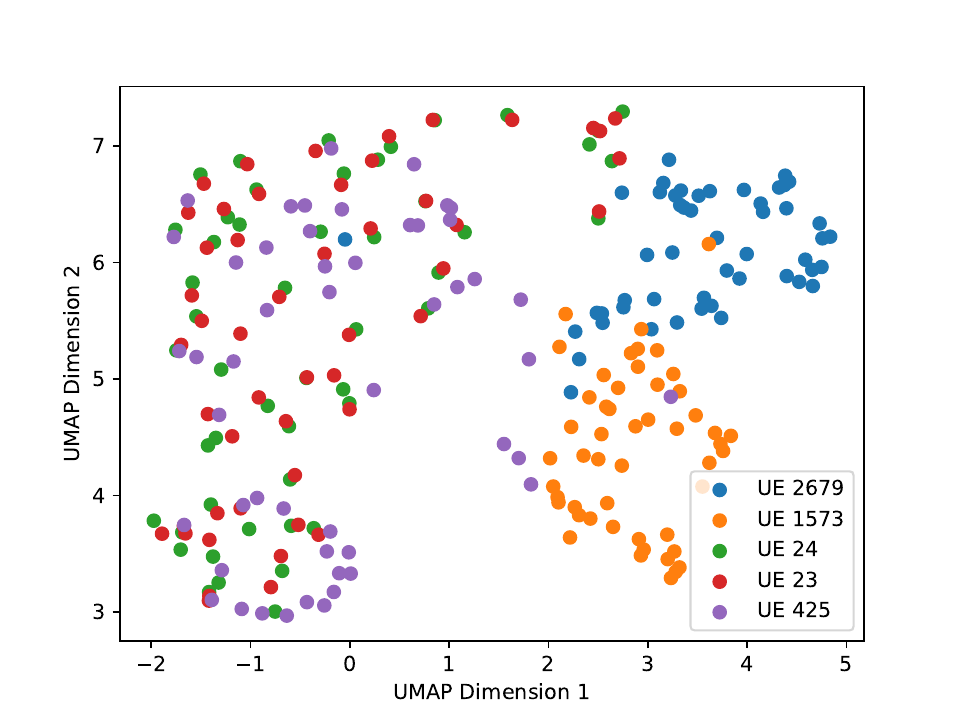}
    \label{fig:1}
  }
  \subfloat[VQ-VAE representation.]
  {
    \includegraphics[width=0.3\linewidth]{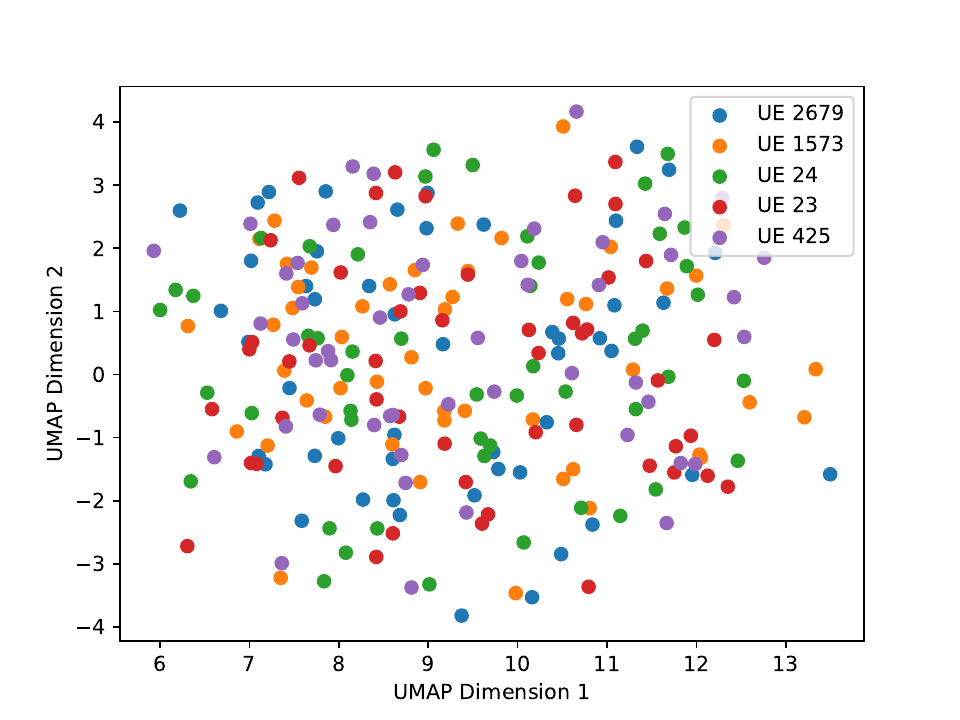}
    \label{fig:2}
  }
  \subfloat[Our representation.]
  {
    \includegraphics[width=0.3\linewidth]{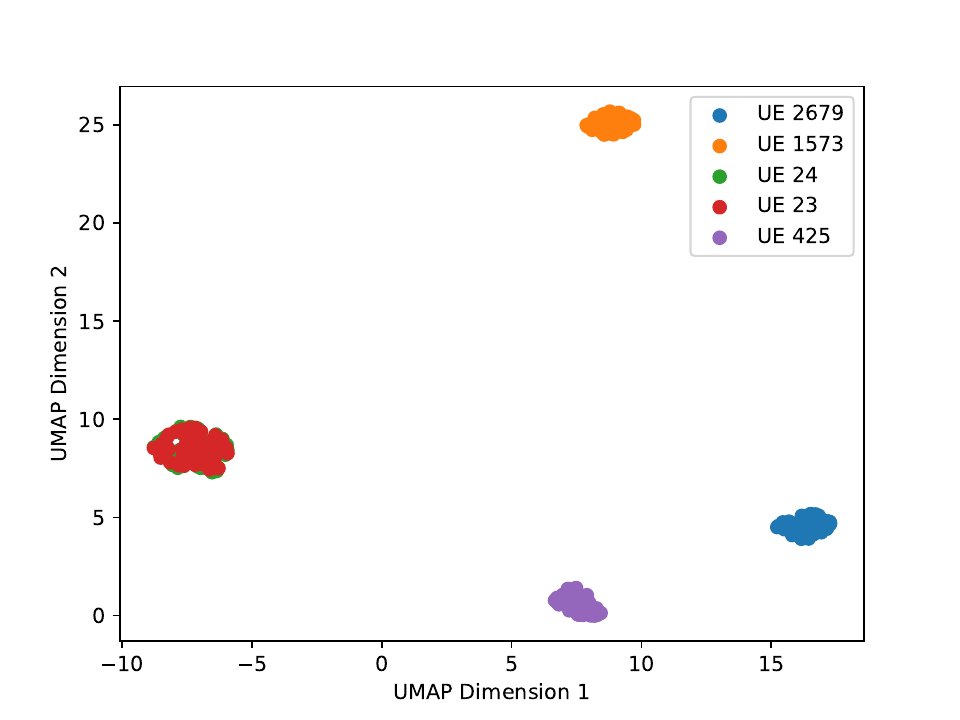}
    \label{fig:3}
  }
  \caption{Visualization of five UEs' channels and  representations based on UMAP.}
  \label{fig:umap}
  \end{minipage}
  \begin{minipage}{0.25\textwidth}
      \centerline{\includegraphics[width=\linewidth]{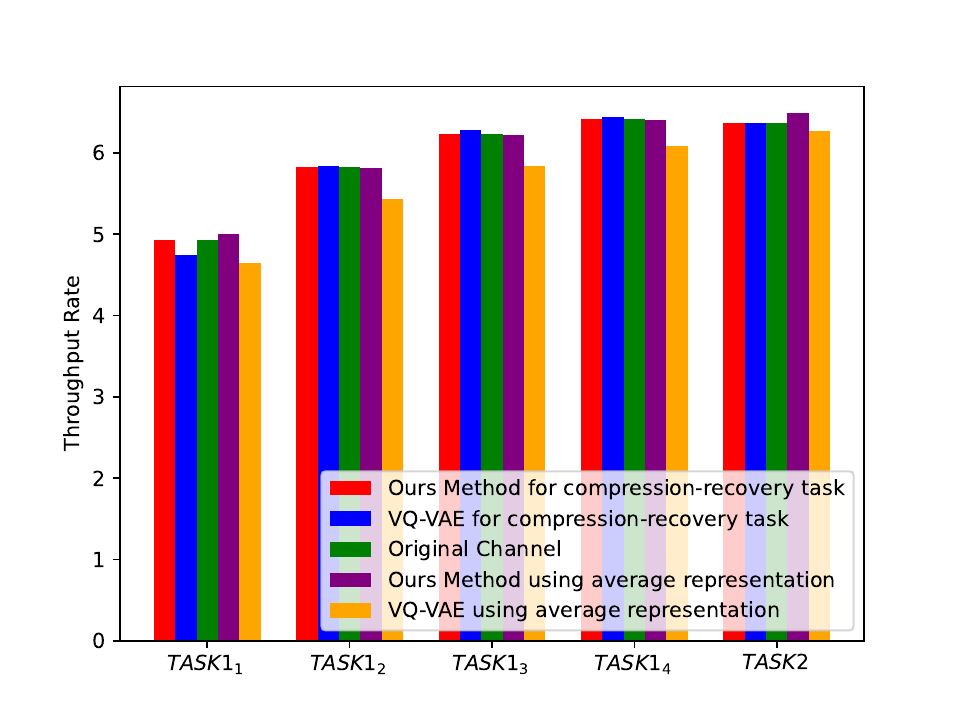}}
\caption{The performance comparison of representation for $\text{TASK1}_{N_{l}}$ and TASK2 between our method and baselines.}
\label{fig:rep_bar}
  \end{minipage}
\end{figure*}
% \begin{figure}[htbp]
% \centerline{\includegraphics[width=0.8\linewidth]{figures/representation_layers_bar_chart.pdf}}
% \caption{The performance comparison of representation based on $\text{TASK1}_{N_{l}}$ and TASK2 between our method and baselines.}
% \label{fig:rep_bar}
% \end{figure}

% \begin{figure*}[htbp]
%     \centering
%     \begin{minipage}{0.32\textwidth}
%         \centering
%         \includegraphics[width=\linewidth]{figures/representation_layers_bar_chart.pdf}
%         \captionsetup{font=footnotesize}
%         \captionof{figure}{The comparison of throughput of $\text{TK1}$ between our method and benchmarks.}
%         \label{fig:re_single_bar}
%     \end{minipage}
%     \hfill
%     \begin{minipage}{0.33\textwidth}
%         \centering
%         \includegraphics[width=\linewidth]{figures/representation_joint_bar_chart.pdf}
%         \captionsetup{font=footnotesize}
%         \captionof{figure}{The comparison of throughput of $\text{TK2}$ between our method and benchmarks.}
%         \label{fig:re_joint_bar}
%     \end{minipage}
%     \hfill
%     \begin{minipage}{0.319\textwidth}
%         \centering
%         \includegraphics[width=\linewidth]{figures/representation_single_CDF.pdf}
%         \captionsetup{font=footnotesize}
%         \captionof{figure}{The CDF of throughput of $\text{TK1}$. }
%         \label{fig}
%     \end{minipage}
% \end{figure*}
\begin{figure*}[htbp]
    \centering
    \begin{minipage}{0.475\textwidth}
        \centering
  \subfloat[CDF.]
  {
    \includegraphics[width=0.49\linewidth]{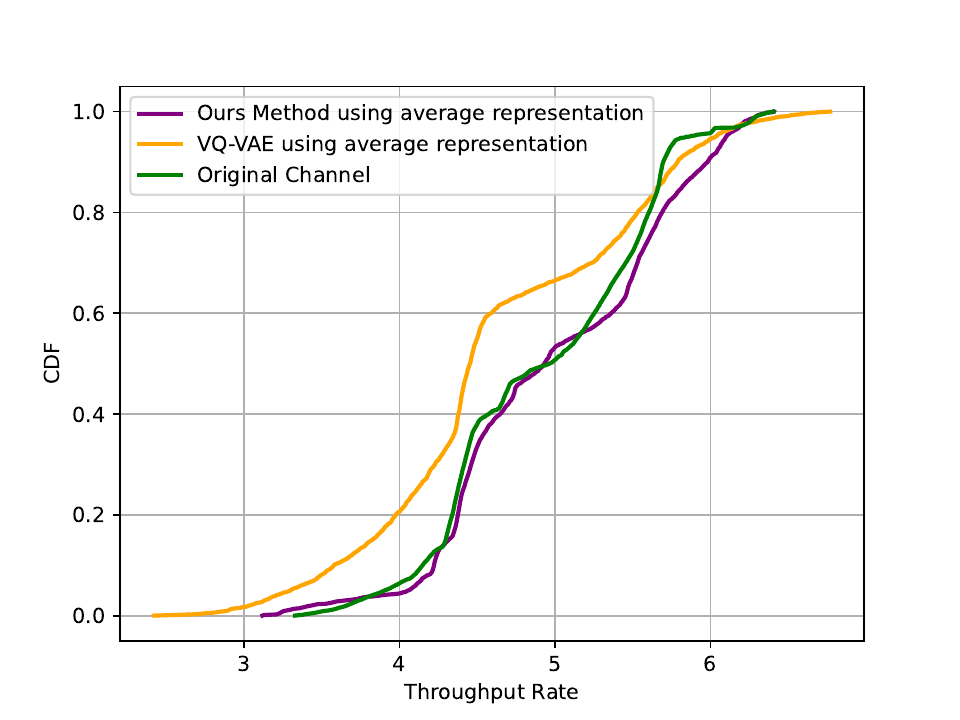}
  }
  \subfloat[Line chart.]
  {
    \includegraphics[width=0.49\linewidth]{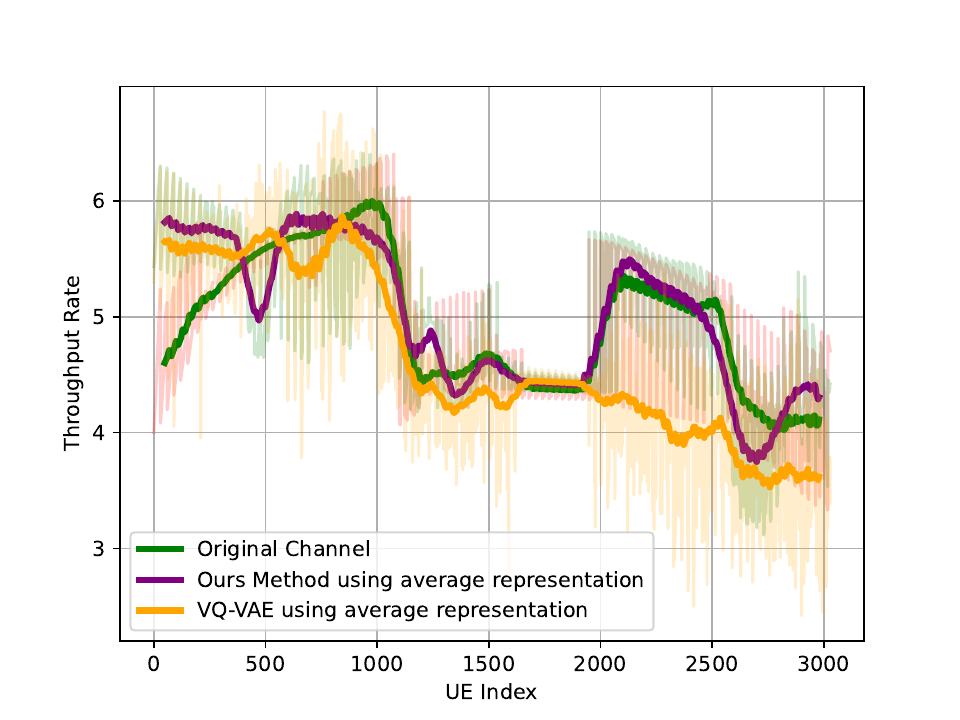}
  }
  \caption{The performance comparison of representation between our method and baselines on $\text{TASK1}_1$ for 3030 UEs.}
  \label{fig:rep_single}
    \end{minipage}
    \hfill
    \begin{minipage}{0.47\textwidth}
    \centering
  \subfloat[CDF.]
  {
    \includegraphics[width=0.49\linewidth]{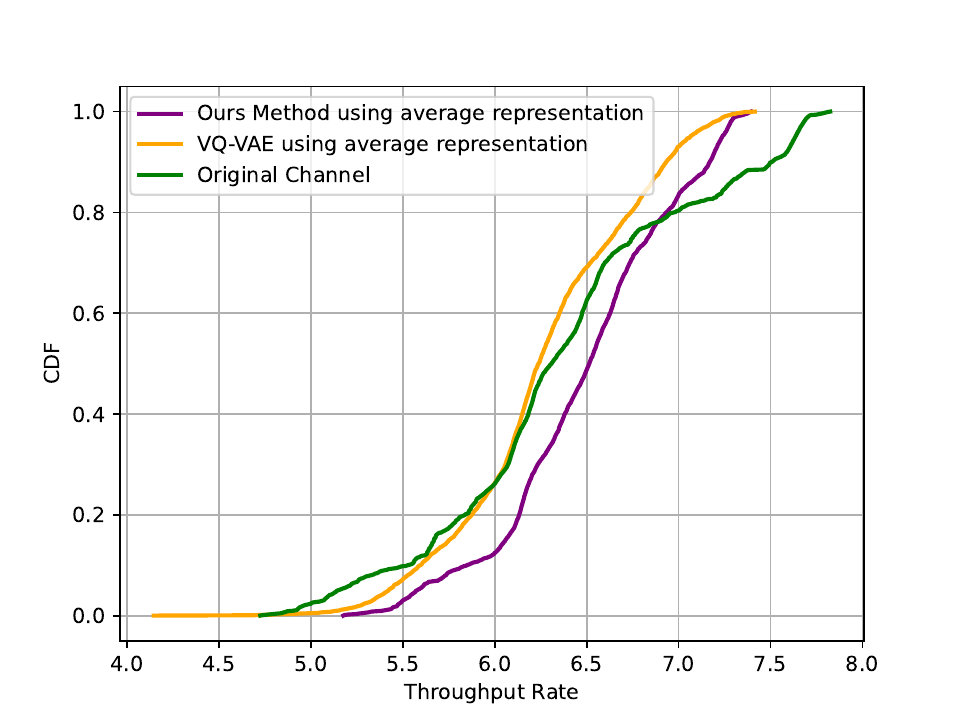}
  }
  \subfloat[Line chart.]
  {
    \includegraphics[width=0.49\linewidth]{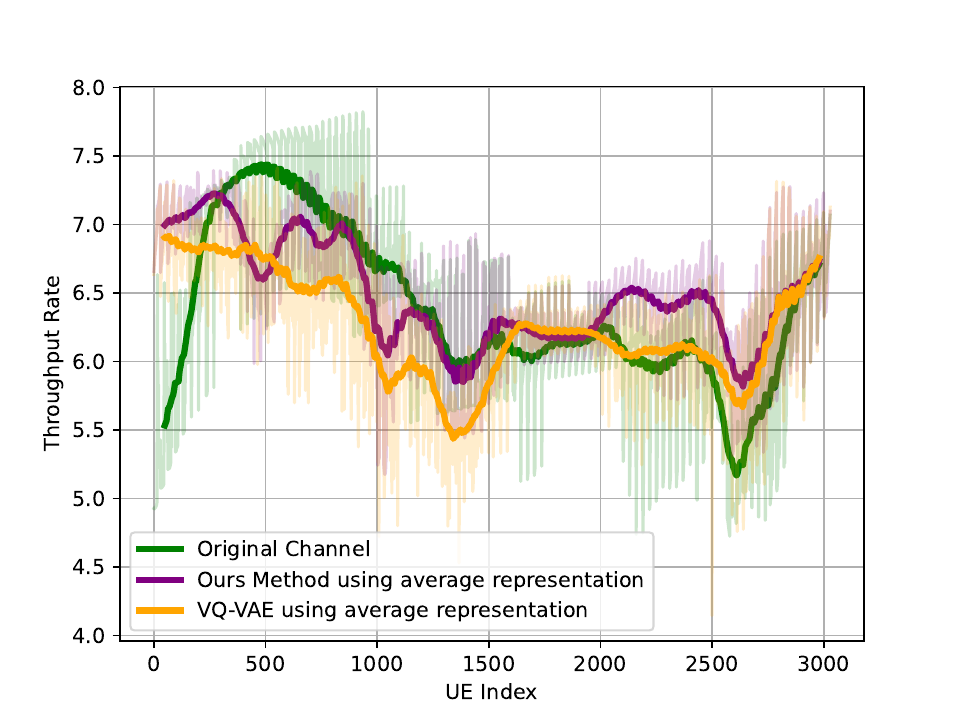}
  }
  \caption{The performance comparison of representation between our method and baselines on $\text{TASK2}$ for 3030 UEs.}
  \label{fig:rep_joint}
    \end{minipage}
\end{figure*}

To verify the performance of proposed method in channel representation, we assess it in terms of the latent space and the geolocation-based MIMO transmission task. The hyperparameters of our method and the VQ-VAE-based method for the encoder and the decoder are shown Table \ref{tb:Hyperparameters encoder}.

In latent space, we utilize the UMAP approach to map the channel and its representation into 2D for visualization, which allows us to qualitatively analyze the quality of the representation. UMAP is a nonlinear dimensionality reduction technique commonly used for visualization and exploring the inherent structures of data. We select five UEs out of 3030, namely UE 2679, 1573, 24, 23, 425, with two adjacent to each other and the remaining three more dispersed. As depicted in Fig. \ref{fig:umap}, the visualization result shows that the channels between the two neighboring UEs are nearly identical, aligning with the expectation that channels at neighboring geolocations should be similar. In contrast, for points in other geolocations, the channel distribution lacks a discernible pattern. The representation derived from the VQ-VAE-based method appears to be almost random, since this approach merely compresses and recovers the channel, which offers little valuable information in the latent space. Conversely, our method effectively identifies similarities between channels of neighboring UEs and differences elsewhere. Theoretically, in contrastive learning, the channels of UE23 and UE24 should be as dissimilar as possible, as they represent distinct samples. However, the chosen value of temperature parameter $\gamma$ in Eq.~\eqref{eq: encoder loss}, results in a very smooth distribution, enabling the network to potentially identify positive samples. This demonstrates that our method captures the inherent nature of the channel, rather than merely compressing channel information.

Then, we validate the performance of our method based on the proposed $\text{TASK1}_{N_{l}}$ and TASK2 metrics. For different methods, we identify the \textit{representative channel} and then calculate the $\text{TASK1}_{N_{l}}$ and TASK2 using the optimal precoding for this channel. The specific results are shown in Fig. \ref{fig:rep_bar}, showing that both the VQ-VAE-based method and our proposed method achieve results nearly identical to those of the original channel on the compression-recovery task, indicating accurate channel recovery by both encoder-decoder pairs. However, when averaging the $N_t$ representations in the latent space and mapping these back to the channel via the decoder, our method demonstrates clear and significant superiority. Averaging the representations serves as a simple form of representation generation, suggesting that our method is robust and can capture essential channel features. This also implies that performing the generation task in the latent space of the proposed method is likely to yield better results.

% \begin{minipage}{\linewidth}
% \begin{minipage}[t]{0.53\linewidth}
% \makeatletter\def\@captype{table}
% % \begin{table}[htbp]
% %     \centering
% %     \caption{Hyperparameters of MoCo Training}
%     \begin{tabular}{c|c}
%         \toprule
%         Parameter  & Value\\
%         \midrule
%         $N_S$& 600\\
%         $\delta_s$      & 1$e^{-4}$+1.65$e^{-5}$s\\
%         $\gamma$      & 5e-3\\
%         $N_{re}$    & 128\\
%         \midrule
%         $d_{model}$& 256\\
%         $L$      & 4\\
%         $\texttt{num\_head}$      & 8\\
%         \bottomrule
    
%     \end{tabular}
%     \caption{Hyperparameters of Diffusion Training}
%     \label{tb:Hyperparameters diff}
% % \end{table}
% \end{minipage}
% \begin{minipage}[t]{0.45\linewidth}
% \makeatletter\def\@captype{table}
% % \begin{table}[htbp]
% %     \centering
% %     \caption{Hyperparameters of VQ-VAE based method}
%     \begin{tabular}{c|c}
%         \toprule
%         Parameter  & Value\\
%         \midrule
%         $d_{model}$& 256\\
%         $L$      & 4\\
%         $\texttt{num\_head}$      & 8\\
%         \bottomrule
        
%     \end{tabular}
%     \caption{Hyperparameters of VQ-VAE based generation method}
%     \label{tb:Hyperparameters VQVAE of gen}
% % \end{table}
% \end{minipage}
% \end{minipage}
Furthermore, we plot the cumulative distribution function (CDF) and line charts for $\text{TASK1}_1$ and TASK2 across 3030 UEs as shown in Fig. \ref{fig:rep_single} and Fig. \ref{fig:rep_joint}, respectively. Since a direct line chart of 3030 UEs does not clearly illustrate the performance differences between methods, we apply window smoothing to the data. The raw data are displayed alongside in lighter colors for comparison. Additionally, since our proposed method and the VQ-VAE-based method for the compression-recovery task perform almost identically to the original channel, they are omitted in these plots. It is evident that after averaging the representations, our method and the original channel alternately outperform each other at times, but both significantly surpass the VQ-VAE-based method.
\subsection{Generation Performance}
\begin{figure*}[htbp] % 使用 figure* 环境来使图片跨过所有栏
  \centering
  
  \begin{minipage}{0.25\textwidth}
        \centering
        \includegraphics[width=\linewidth]{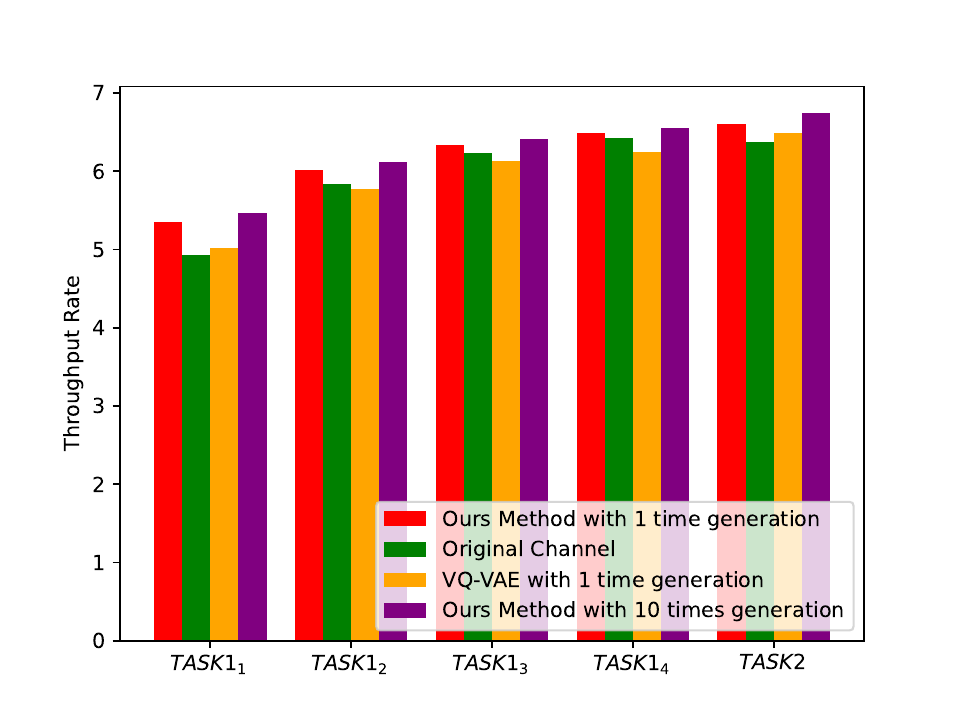}
        \caption{The performance comparison of generation for $\text{TASK1}_{N_{l}}$ and TASK2 between our method and baselines.}
\label{fig:gen_bar}
    \end{minipage}
    \hfill
    \begin{minipage}{0.47\textwidth}
        \centering
        \subfloat[Representations.]
  {
    \includegraphics[width=0.48\linewidth]{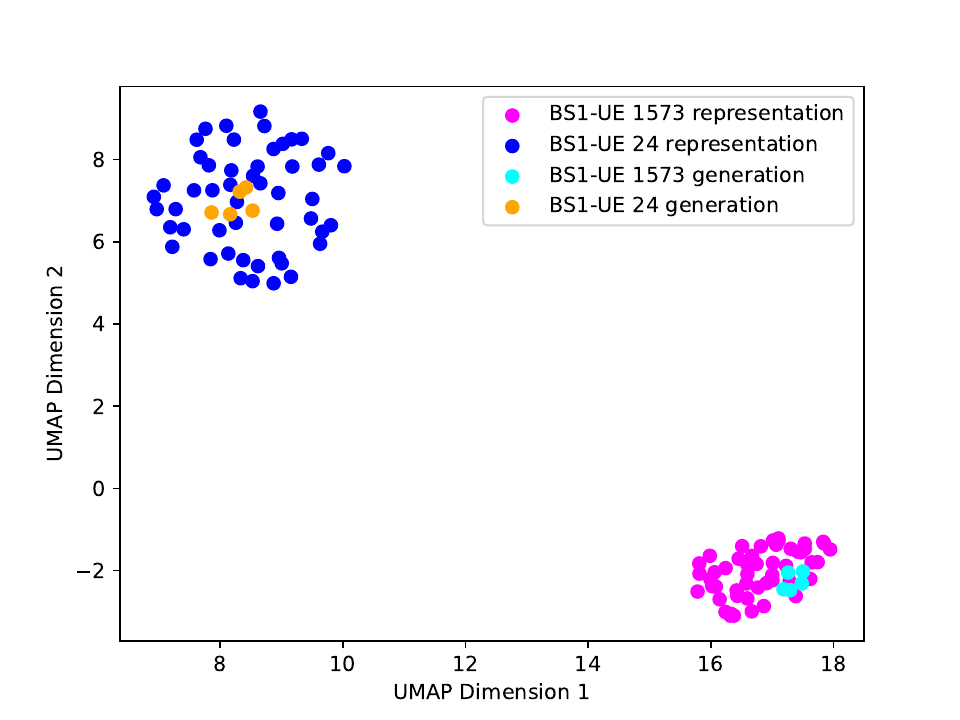}
    \label{fig:gen_1}
  }
  \subfloat[Channels.]
  {
    \includegraphics[width=0.48\linewidth]{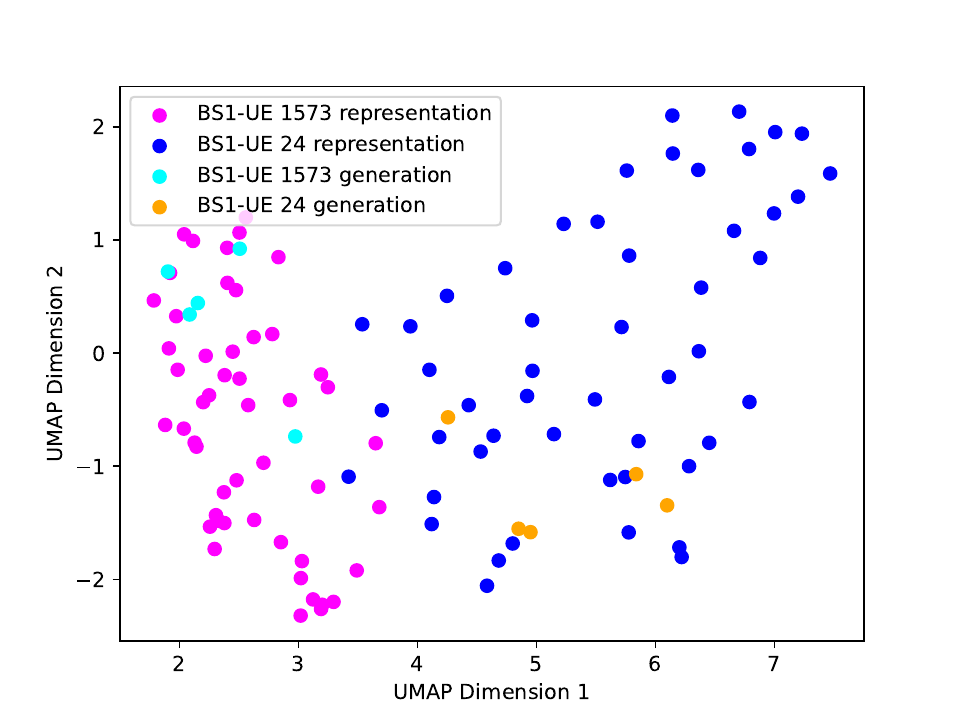}
    \label{fig:gen_2}
  }
  \caption{Visualization of generation for two specific UEs.}
  \label{fig:gen_umap}
    \end{minipage}
    \hfill
    \begin{minipage}{0.24\textwidth}
        \centering
        \includegraphics[width=\linewidth]{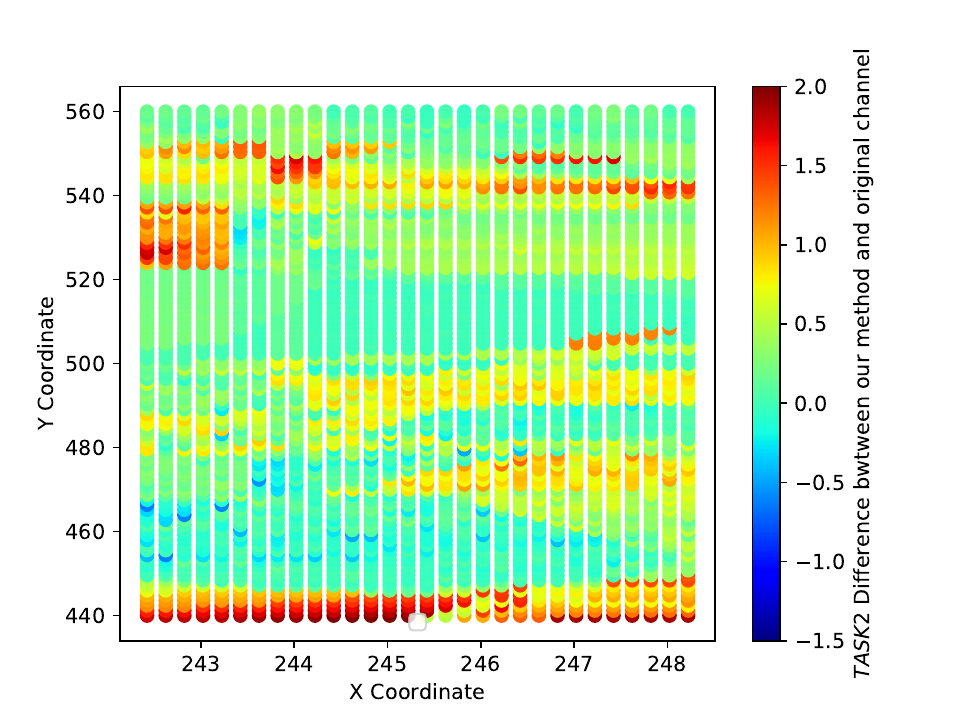}
        \caption{Heatmap of $\text{TASK1}_1$ gap between our method and original channel.}
\label{fig:gen_bar}
    \end{minipage}
\end{figure*}

\begin{figure*}[htbp]
    \centering
    \begin{minipage}{0.47\textwidth}
        \centering
  \subfloat[CDF.]
  {
    \includegraphics[width=0.49\linewidth]{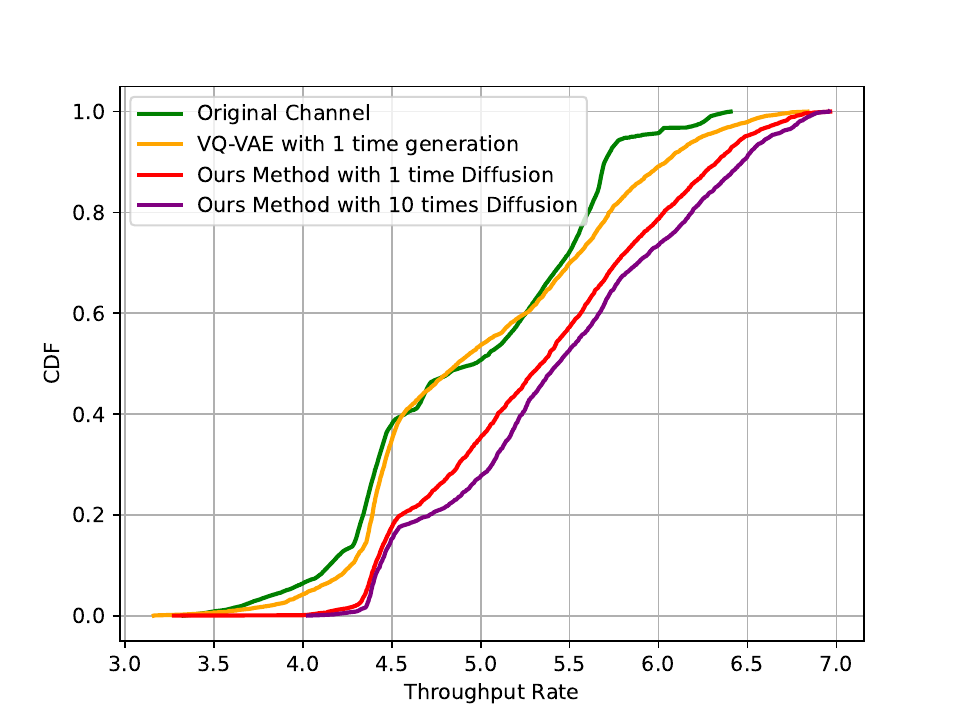}
    \label{fig:1}
  }
  \subfloat[Line chart.]
  {
    \includegraphics[width=0.49\linewidth]{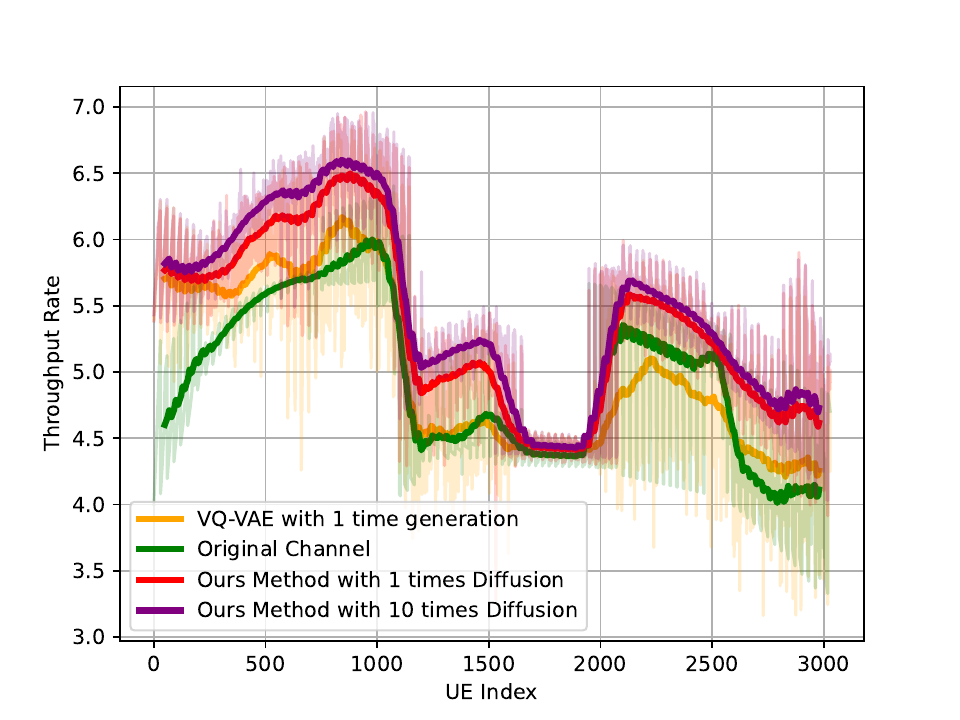}
    \label{fig:2}
  }
  \caption{The performance comparison of generation  between our method and baselines on $\text{TASK1}_1$ for 3030 UEs.}
  \label{fig:gen_TK1}
    \end{minipage}
    \hfill
    \begin{minipage}{0.47\textwidth}
    \centering
    \subfloat[CDF.]
  {
    \includegraphics[width=0.49\linewidth]{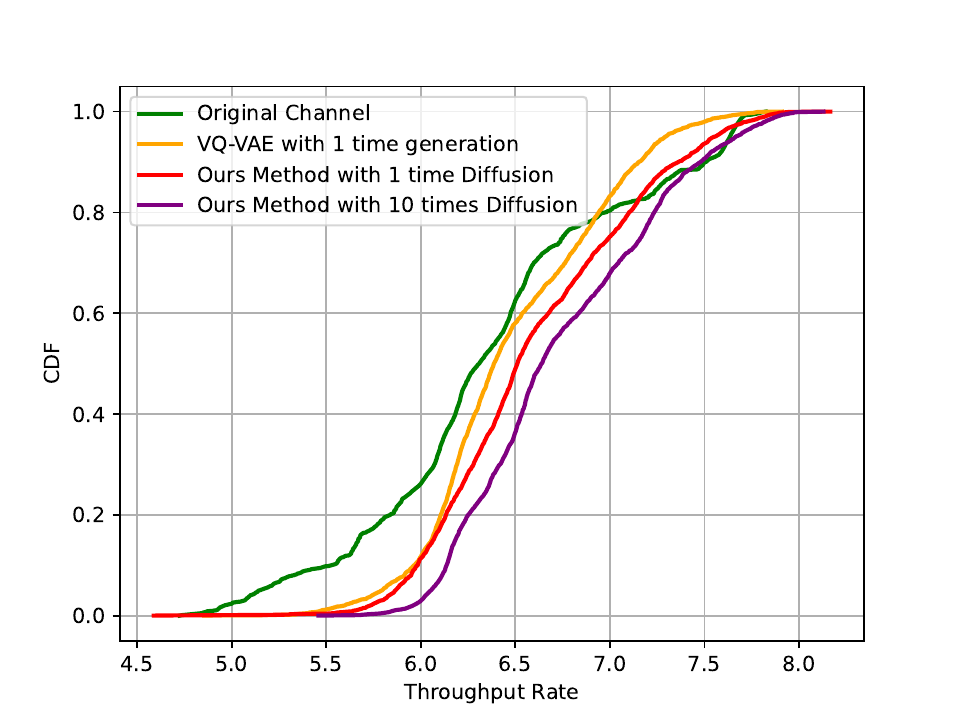}
    \label{fig:1}
  }
  \subfloat[Line chart.]
  {
    \includegraphics[width=0.49\linewidth]{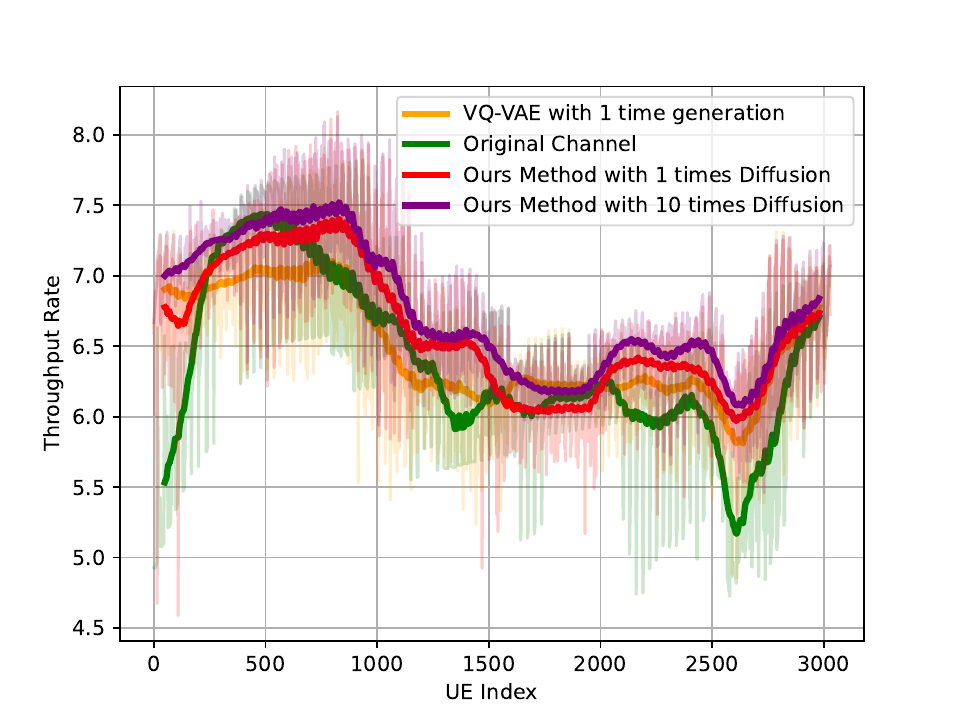}
    \label{fig:2}
  }
  \caption{The performance comparison of generation between our method and baselines on $\text{TASK2}$ for 3030 UEs.}
  \label{fig:gen_TK2}
    \end{minipage}
\end{figure*}
\begin{table}[!htbp]
	\caption{Hyperparameters of the Generator for our Method and the VQ-VAE-based Method}	\centering
	\begin{tabular}{c|c}
        \toprule
        Parameter  & Value\\
        \midrule
        $N_S$& 600\\
        $\delta_s$      & 1$e^{-4}$+1.65$e^{-5}$s\\
        Learning rate      & 5e-3\\
        \midrule
        $d_{model}$& 256\\
        $L$      & 4\\
        Number of multi-head      & 8\\
        \bottomrule
    \end{tabular}
    \label{tb:Hyperparameters generator}
\end{table}
We similarly conduct performance validation in the latent space and the geolocation-based MIMO transmission task for generation results. The hyperparameters of the conditional diffusion model in our method and the GPT-like model in the VQ-VAE-based method are shown Table \ref{tb:Hyperparameters generator}.

In the latent space, we utilize UMAP to visualize both the generated and original channel representations. We then remap the generated representation back to the channel using the decoder and subsequently visualize the original and generated channels using UMAP. It is observable in Fig. \ref{fig:gen_umap} that both the generated representation and channel exhibit similarity to the original channel, indicating that the diffusion model effectively learns the distribution characteristics of the representation and can generate new, plausible representations and channels.

Furthermore, we evaluate the performance of our generated channels, comparing them with the original channel traversal and the VQ-VAE-based generation methods. The performance scores for these methods on $\text{TASK1}_1$ and TASK2 are shown in Fig. \ref{fig:gen_bar}, Fig. \ref{fig:gen_TK2} and Fig. \ref{fig:gen_TK1}, respectively. We generate representations 1 and 10 times, respectively, recover these representations as channels, and select the best channel as the \textit{representative channel}. It is evident that our generated channel effectively represents the inherent nature learned during training, resulting in superior performance across all tests compared to other baselines. Additionally, since generating channels via diffusion is virtually cost-free, we can repeatedly generate channels and select the best-performing one to enhance performance further. Our method outperforms the two baselines with just 1 time generation and continues to improve with up to 10 times generations, demonstrating its effectiveness, enhancing geolocation-based MIMO transmission at all locations.

It should be noted that in the $\text{TASK1}_{N_l}$ task, the relative improvement of our method compared to the baseline diminishes as $N_{L}$ increases, which can be seen from Fig. 16. That is, as employing using more layers in MIMO transmission necessitates that our \textit{representative channel} can closely approximate the actual channel. Given that the number of layers represents the dominant paths, favorable outcomes can be achieved if $N_{l} = 1$ and our \textit{representative channel} mirrors the main path of the real channel. However, as $N_{l}$ increases, aligning with additional paths becomes more challenging, leading to smaller gains of the proposed method to baselines.

\section{Conclusion}
In this paper, we have proposed a self-supervised learning method for MIMO-OFDM channel representation and generation. We have designed a pretext task that leverages the temporal and spatial domain correlations of the channel to obtain a representation in hidden space through self-supervised learning. To evaluate the quality of this representation, we have proposed geolocation-based MIMO transmission as a downstream task. A conditional diffusion generator, trained in the latent space, and a decoder, are employed to execute this downstream task. Simulation results using public ray-tracing-based channel datasets demonstrate that our method performs comparably to other existing methods when using only the encoder and the decoder and surpasses other existing methods after employing the generator. In the future, we will further explore methods for acquiring representations and investigate their applications in other prevalent communication tasks.

\section*{Acknowledgment}
This work was supported in part by the National Natural Science Foundation Original Exploration Project of China under Grant 62250004, the Natural Science Foundation of China (NSFC) under Grant 62001259, 62271244, the Natural Science Fund for Distinguished Young Scholars of Jiangsu Province under Grant BK20220067, the National R\&D Program of China (2020YFB1807503), China National Postdoctoral Program for Innovative Talents under Grant BX20230163, China Postdoctoral Science Foundation under Grant 2023M731832 and The Major Key Project of PCL.

\begin{appendices}
\normalfont
\section{Proof of \textbf{Theorem 1}}
The spectral efficiency at subcarrier $k$ and time $t$ is formulated as follows:
\begin{equation}
    R_{k,t} \!\!\coloneqq \!\!\log_2\left(\!1+\frac{\|\mathbf{H}^{b_1,u,t,k}\mathbf{W}^{b_1,u,t,k}\!\!+\!\mathbf{H}^{b_2,u,t,k}\mathbf{W}^{b_2,u,t,k}\|_2^2}{\sigma_n^2}\right)_{.}
\end{equation}
Denote $\mathbf{W}^{b_1,b_2,u,t,k} = \left[\mathbf{W}^{b_1,u,t,k}, \mathbf{W}^{b_2,u,t,k}\right]^T\in \mathbb{C}^{2N_T\times1}$, we have
\begin{equation}
\begin{aligned}
   R_{k,t}&=\log_2\left(1+\frac{1}{\sigma_n^2} (\mathbf{W}^{b_1,b_2,u,t,k})^H (\mathbf{H}^{b_1, b_2,u,t,k})^H\right.\\
   &~~~~~~~~~~~~\left.\mathbf{H}^{b_1, b_2,u,t,k} \mathbf{W}^{b_1,b_2,u,t,k}\vphantom{\frac{1}{\sigma_n^2}}\right)
   \\
   &\overset{(1)}{=}\log_2\left|\mathbf{I}+\frac{1}{\sigma_n^2} \mathbf{H}^{b_1, b_2,u,t,k} \mathbf{W}^{b_1,b_2,u,t,k}\right.\\
   &~~~~~~~~~~~~\left.(\mathbf{W}^{b_1,b_2,u,t,k})^H (\mathbf{H}^{b_1, b_2,u,t,k})^H\vphantom{\frac{1}{\sigma_n^2}}\right|.
\end{aligned}
\end{equation}
where $\overset{(1)}{=}$ is based on $|I+AB|=|I+BA|$. Denote $\mathbf{Q}^{b_1,b_2,u,t,k}=\mathbf{W}^{b_1,b_2,u,t,k}(\mathbf{W}^{b_1,b_2,u,t,k})^H\in \mathbb{C}^{2N_T\times 2N_T}$, we have
\begin{equation}
    \begin{aligned}
        R_{k,t} &= \log_2\left|\mathbf{I}+\frac{1}{\sigma_n^2}\mathbf{H}^{b_1, b_2,u,t,k}\mathbf{Q}^{b_1,b_2,u,t,k}(\mathbf{H}^{b_1, b_2,u,t,k})^H\right|
        \\
        &\overset{(1)}{=}\log_2\left|\mathbf{I}+\frac{1}{\sigma_n^2}\mathbf{Q}^{b_1,b_2,u,t,k}(\mathbf{H}^{b_1, b_2,u,t,k})^H\mathbf{H}^{b_1, b_2,u,t,k}\right|
        \\
        &\overset{(2)}{=}\log_2\left|\mathbf{I}+\frac{1}{\sigma_n^2}\mathbf{Q}^{b_1,b_2,u,t,k}\right.\\&~~~~~~~~~~~~~~~~~\left.(\mathbf{U}^{b_1, b_2,u,t,k})^H \Sigma^{b_1, b_2,u,t,k}\mathbf{U}^{b_1, b_2,u,t,k}\vphantom{\frac{1}{\sigma_n^2}}\right|
        \\&\overset{(1)}{=}\log_2\left|\mathbf{I}+\frac{1}{\sigma_n^2}(\Sigma^{b_1, b_2,u,t,k})^{\frac{1}{2}}\mathbf{U}^{b_1, b_2,u,t,k}\mathbf{Q}^{b_1,b_2,u,t,k}\right.\\&~~~~~~~~~~~~~~~~~\left.(\mathbf{U}^{b_1, b_2,u,t,k})^H (\Sigma^{b_1, b_2,u,t,k})^{\frac{1}{2}}\vphantom{\frac{1}{\sigma_n^2}}\right|.
    \end{aligned}
\end{equation}
where $\overset{(2)}{=}$ represents eigenvalue decomposition on the semi-definite matrix $(\mathbf{H}^{b_1, b_2,u,t,k})^H\mathbf{H}^{b_1, b_2,u,t,k}$, and we can obtain that $(\mathbf{H}^{b_1, b_2,u,t,k})^H\mathbf{H}^{b_1, b_2,u,t,k}=(\mathbf{U}^{b_1, b_2,u,t,k})^H\Sigma^{b_1,b_2,u,t,k}\mathbf{U}^{b_1, b_2,u,t,k}$. Let $\hat{\mathbf{Q}}^{b_1,b_2,u,t,k}=\mathbf{U}^{b_1, b_2,u,t,k}\mathbf{Q}^{b_1,b_2,u,t,k}(\mathbf{U}^{b_1, b_2,u,t,k})^H \in\mathbf{C}^{2N_T\times 2N_T}$, we have
\begin{equation}
    R_{k,t} = \log_2\left|\mathbf{I}+\frac{1}{\sigma_n^2}(\Sigma^{b_1, b_2,u,t,k})^{\frac{1}{2}}\hat{\mathbf{Q}}^{b_1,b_2,u,t,k}(\Sigma^{b_1, b_2,u,t,k})^{\frac{1}{2}}\right|.
\end{equation}
 According to the power constraints, we have
\begin{equation}
\begin{aligned}
    &\|\hat{\mathbf{Q}}^{b_1,b_2,u,t,k}\|_F = \sqrt{\operatorname{tr}\left((\hat{\mathbf{Q}}^{b_1,b_2,u,t,k})^H\hat{\mathbf{Q}}^{b_1,b_2,u,t,k}\right)}
    \\&=\sqrt{
    \begin{aligned}\operatorname{tr}\left(\mathbf{U}^{b_1, b_2,u,t,k}\mathbf{Q}^{b_1,b_2,u,t,k}(\mathbf{U}^{b_1, b_2,u,t,k})^H\right.\\\left.\mathbf{U}^{b_1, b_2,u,t,k}(\mathbf{Q}^{b_1,b_2,u,t,k})^H(\mathbf{U}^{b_1, b_2,u,t,k})^H\right)
    \end{aligned}
    }
    \\&\overset{(3)}{=}\sqrt{
    \begin{aligned}\operatorname{tr}\left(\mathbf{U}^{b_1, b_2,u,t,k}\mathbf{W}^{b_1,b_2,u,t,k}(\mathbf{W}^{b_1,b_2,u,t,k})^H\right.\\\left.\mathbf{W}^{b_1,b_2,u,t,k}(\mathbf{W}^{b_1,b_2,u,t,k})^H(\mathbf{U}^{b_1, b_2,u,t,k})^H\right)
    \end{aligned}
    }
    \\&=\sqrt{
    \begin{aligned}\operatorname{tr}\left((\mathbf{W}^{b_1,b_2,u,t,k})^H\mathbf{W}^{b_1,b_2,u,t,k}\right.\\\left.(\mathbf{W}^{b_1,b_2,u,t,k})^H\mathbf{W}^{b_1,b_2,u,t,k}\right)
    \end{aligned}}
    % \\&=\sqrt{\operatorname{tr}\left((\mathbf{W}^{b_1,b_2,u,t,k})^H\mathbf{W}^{b_1,b_2,u,t,k}(\mathbf{W}^{b_1,b_2,u,t,k})^H\mathbf{W}^{b_1,b_2,u,t,k}\right)}
    \\&=\left\|(\mathbf{W}^{b_1,b_2,u,t,k})^H\mathbf{W}^{b_1,b_2,u,t,k}\right\|_F
    \\&=\left\|\mathbf{W}^{b_1,u,t,k}\right\|_F + \left\|\mathbf{W}^{b_2,u,t,k}\right\|_F
    \\&=P^{b_1,u,t,k} + P^{b_2,u,t,k} \leq 2N_k.
\end{aligned}
\end{equation}
where $\operatorname{tr}$ represents the trace operation of a matrix and $\overset{(3)}{=}$ is based on $\operatorname{tr}(AB)=\operatorname{tr}(BA)$. We can see that $\|\hat{\mathbf{Q}}^{b_1,b_2,u,t,k}\|_F$ is limited. Therefore, only by making the non-diagonal elements of $\hat{\mathbf{Q}}^{b_1,b_2,u,t,k}$ equal 0 can the diagonal element of $\hat{\mathbf{Q}}^{b_1,b_2,u,t,k}$ be maximized.
In other words, $\left|\mathbf{I}+\frac{1}{\sigma_n^2}(\Sigma^{b_1, b_2,u,t,k})^{\frac{1}{2}}\hat{\mathbf{Q}}^{b_1,b_2,u,t,k}(\Sigma^{b_1, b_2,u,t,k})^{\frac{1}{2}}\right|$ can be maximized if $\hat{\mathbf{Q}}^{b_1,b_2,u,t,k}$ is a diagonal matrix. If $\hat{\mathbf{Q}}^{b_1,b_2,u,t,k}$ are all diagonal matrices for $\forall k \in N_k$, we can derive that
\begin{equation}
\begin{aligned}
    \sum_{k=1}^{N_k}R_{k,t}&\overset{(1)}{=}\sum_{k=1}^{N_k}\log_2\left|\mathbf{I}+\frac{1}{\sigma_n^2}\hat{\mathbf{Q}}^{b_1,b_2,u,k}\Sigma^{b_1, b_2,u,t,k}\right|
    \\&=\log_2\prod_{k=1}^{N_k}\left|\mathbf{I}+\frac{1}{\sigma_n^2}\hat{\mathbf{Q}}^{b_1,b_2,u,k}\Sigma^{b_1, b_2,u,t,k}\right|
    \\&=\log_2\left|\mathbf{I}+\frac{1}{\sigma_n^2}\operatorname{diag}\left(\left[\hat{\mathbf{Q}}^{b_1,b_2,u,1}\Sigma^{b_1, b_2,u,t,1}, 
    \right.\right.\right.\\&~~~~~~~~~~\left.\left.\left.\ldots, \hat{\mathbf{Q}}^{b_1,b_2,u,N_k}\Sigma^{b_1, b_2,u,t,N_k}
 \right]\right)\vphantom{\frac{1}{\sigma_n^2}} \right|,
\end{aligned}
\end{equation}
where $\operatorname{diag}$ denotes the diagonalization operation. Assume that the diagonal elements of $\Sigma^{b_1, b_2,u,t,k}$ are arranged from biggest to smallest, and the biggest element is $(\lambda^{b_1,b_2,u,t,k})^2$.
Thus, to maximize $\sum_{k=1}^{N_k}R_{k,t}$, the optimal matrix $\Sigma_{*}^{b_1,b_2,u,t,k}$ is
\begin{equation}
\begin{aligned}
\!\!\hat{\mathbf{Q}}^{b_1,b_2,u,k}=\Sigma_{*}^{b_1,b_2,u,t,k}=\!\!\left[
   \begin{array}{ccccc}

       P^{b_1,u,k}+P^{b_2,u,k} \!\!\!& 0 & \cdots & 0   \\
      0 \!\!\!& 0 &\cdots & 0 \\
       \vdots \!\!\!&  \vdots   & & \vdots \\
      0 \!\!\!& 0  &  \cdots & 0\\
   \end{array}
 \right]_{.}
\end{aligned}
\end{equation}
Therefore, we have:
\begin{equation*}
\begin{aligned}
    &\sum_{k=1}^{N_k}R_{k,t} = \log_2\left|\mathbf{I}+\frac{1}{\sigma_n^2}\operatorname{diag}\left(\left[\Sigma_{*}^{b_1,b_2,u,t,1}\Sigma^{b_1, b_2,u,t,1},\right.\right.\right.\\&~~~~~~~~~~\left.\left.\left.\ldots, \Sigma_{*}^{b_1,b_2,u,t,N_k}\Sigma^{b_1, b_2,u,t,N_k}
 \right]\right)\vphantom{\frac{1}{\sigma_n^2}} \right|
 \end{aligned}
\end{equation*}
\begin{equation}
\begin{aligned}
 &=\log_2\prod_{k=1}^{N_k}\left(1+\frac{1}{\sigma_n^2}(\lambda^{b_1,b_2,u,t,k})^2\left(P^{b_1,u,k}+P^{b_2,u,k}\right)\right)
\\&=\sum_{k=1}^{N_k}\log_2\left(1+\frac{1}{\sigma_n^2}(\lambda^{b_1,b_2,u,t,k})^2\left(P^{b_1,u,k}+P^{b_2,u,k}\right)\right).
\end{aligned}
\end{equation}
In summary, the optimization problem is transformed into Eq.~\eqref{task2 convert}.
\end{appendices}

\nocite{*}
\vspace*{-10mm}
\bibliographystyle{IEEEtran}\Huge
\bibliography{reference}

\end{document}